\documentclass{emulateapj}

\usepackage{graphics} 
\usepackage{epsfig}
\usepackage{natbib}
\usepackage{rotating} 
\shorttitle{Protoplanetary Disk H$_{2}$ Survey}
\shortauthors{France et al.}
\begin{document}
\title{A {\it Hubble Space Telescope} Survey of H$_{2}$ Emission in the Circumstellar Environments of Young Stars\altaffilmark{*}}
\author{
Kevin France\altaffilmark{1},
Eric Schindhelm\altaffilmark{1,2}, 
Gregory J. Herczeg\altaffilmark{3,4}
Alexander Brown\altaffilmark{1}, 
Herv{\'e} Abgrall\altaffilmark{5},
Richard D. Alexander\altaffilmark{6}, 
Edwin A. Bergin\altaffilmark{7},
Joanna M. Brown\altaffilmark{8},
Jeffrey L. Linsky\altaffilmark{9}, 
Evelyne Roueff\altaffilmark{5}, and
Hao Yang\altaffilmark{10} 
}
\altaffiltext{*}{Based on observations made with the NASA/ESA $Hubble$~$Space$~$Telescope$, obtained from the data archive at the Space Telescope Science Institute. STScI is operated by the Association of Universities for Research in Astronomy, Inc. under NASA contract NAS 5-26555.}
\altaffiltext{1}{Center for Astrophysics and Space Astronomy, University of Colorado, 389 UCB, Boulder, CO 80309, USA; kevin.france@colorado.edu}
\altaffiltext{2}{Current Addesss: Southwest Research Institute, 1050 Walnut Street, Suite 300, Boulder, CO 80302, USA}
\altaffiltext{3}{Max-Planck-Institut f\"{u}r extraterrestriche Physik, Postfach 1312, 85741 Garching, Germany}
\altaffiltext{4}{Current Addesss: Kavli Institute for Astronomy and Astrophysics, Peking University, Beijing 100871, China}
\altaffiltext{5}{LUTH and UMR 8102 du CNRS, Observatoire de Paris, Section de Meudon, Place J. Janssen, 92195 Meudon, France}
\altaffiltext{6}{Department of Physics \& Astronomy, University of Leicester, Leicester, LE1 7RH, UK}
\altaffiltext{7}{Department of Astronomy, University of Michigan, 830 Dennison Building, 500 Church Street, Ann Arbor, MI 48109, USA}
\altaffiltext{8}{Harvard-Smithsonian Center for Astrophysics, 60 Garden Street, MS-78, Cambridge, MA 02138, USA}
\altaffiltext{9}{JILA, University of Colorado and NIST, 440 UCB, Boulder, CO 80309}
\altaffiltext{10}{Institute of Astrophysics, Central China Normal University, Wuhan, Hubei, 430079, China}

\begin{abstract}
The formation timescale and final architecture of exoplanetary systems are closely related to the properties of the molecular disks from which they form. Observations of the spatial distribution and lifetime of the molecular gas at planet-forming radii ($a$~$<$~10~AU) are important for understanding the formation and evolution of exoplanetary systems. Towards this end, we present the largest spectrally resolved survey of H$_{2}$ emission around low-mass pre-main sequence stars compiled to date. We use a combination of new and archival far-ultraviolet spectra from the COS and STIS instruments on the {\it Hubble Space Telescope} to sample 34 T Tauri stars (27 actively accreting CTTSs and 7 non-accreting WTTSs) with ages ranging from 
$\sim$~1~--~10~Myr. 
We observe fluorescent H$_{2}$ emission, excited by Ly$\alpha$ photons, in 100\% of the accreting sources, including all of the transitional disks in our sample (CS Cha, DM Tau, GM Aur, UX Tau A, LkCa15, HD 135344B and TW Hya). 
The spatial distribution of the emitting gas is inferred from spectrally resolved H$_{2}$ line profiles. Some of the emitting gas is produced in outflowing material, but the majority of H$_{2}$ emission appears to originate in a rotating disk. For the disk-dominated targets, the H$_{2}$ emission originates predominately at $a$~$\lesssim$~3~AU. The emission line-widths and inner molecular radii are found to be roughly consistent with those measured from mid-IR CO spectra. 
\end{abstract}

\clearpage

\section{Introduction}

The lifetime, spatial distribution, and composition of gas and dust in the inner $\sim$~10 AU of young (age $\lesssim$~10 Myr) circumstellar disks are important components for understanding of the formation and evolution of extrasolar planetary systems. The formation of giant planet cores and their accretion of gaseous envelopes occurs on timescales similar to the lifetimes of the disks around Classical T Tauri Stars (CTTSs; 10$^{6}$~--~10$^{7}$ yr). The cores of giant planets are thought to be comprised of coagulations of dust grains~\citep{hayashi85}, and the majority of observational work on the lifetime of inner disks has come from photometric and spectroscopic studies of their dust~\citep{haisch01,hernandez07,wyatt08}. Dust in protoplanetary disks is observed as mid- and far-IR excess flux produced by warm grains (e.g., Furlan et al. 2006, 2009; Evans et al. 2009; Luhman et al. 2010).\nocite{furlan06,furlan09,evans09,luhman10} The IR spectral energy distributions (SEDs) of protoplanetary disks are sensitive to the radial distribution of dust in the disk; this dependence has led to the discovery of a class of ``transitional'' systems, whose SEDs indicate that gaps of a few tenths to tens of AU have been opened in their inner disks~(Strom et al. 1989; Calvet et al. 2002, 2005; Espaillat et al. 2007; and see the review by Williams \& Cieza 2011).\nocite{strom89,calvet02,calvet05,williams11} The physical process by which the inner disk is cleared is not yet established. Possible mechanisms including photoevaporation~\citep{alexander06,gorti09} and dynamical clearing by exoplanetary systems~\citep{rice03,dodson11}, possibly aided by the magnetorotational instability~\citep{chiang07}, can reproduce certain transitional disk observations.~\nocite{hernandez07,chiang07}

The lifetime and spatial extent of the gas disk determine the final mass of giant planets~\citep{ida04} and the final architecture of an exoplanetary system, as disk gas regulates type-II planetary migration~\citep{ward97,armitage02,trilling02}. Because the migration timescale is sensitive to the specifics of the disk surface density distribution and dissipation timescale~\citep{armitage07}, observations of gas-rich systems with age~$\leq$~10 Myr can provide important constraints on models of the evolution of exoplanetary systems.
Significant observational effort has been devoted to the study of inner disk gas in recent years, including ground-based mid-IR spectroscopy of CO and [\ion{Ne}{2}] \citep{najita03,pascucci07,herczeg07,najita09,bast11,sacco12}, 
spectroastrometric observations of CO~\citep{pontoppidan08,pontoppidan11} and [\ion{Ne}{2}]~\citep{pascucci11}, and $Spitzer$-IRS observations of H$_{2}$O and organics~\citep{carr08,salyk08,salyk11a,carr11}. There is growing evidence that remnant gas disk survival is common inside the dust hole in transitional disks (e.g., Salyk et al. 2009, 2011), suggesting that planetary migration may continue after the dust disk has dispersed. \nocite{salyk09,salyk11} In transitional systems with minimal inner disk dust, observations of active accretion also provide indirect evidence for the presence of a remnant gas disk~\citep{najita07b,kim09,merin10,fedele10}.

Many previous studies of inner disk gas have used trace species to infer the presence of molecular hydrogen (H$_{2}$), the primary constituent of protoplanetary disks and gas giant planets. The homonuclear
nature of H$_{2}$ means that rovibrational transitions are dipole forbidden, with weak quadrupole transitions that have large energy spacings. This makes direct detection of H$_{2}$ challenging at near- and mid-IR wavelengths (Pascucci et al. 2006; Carmona et al. 2008; but see also Bary et al. 2008).~\nocite{pascucci06,carmona08,bary08} However, H$_{2}$ can be observed in the far-UV (912~--~1650~\AA) bandpass, where the strong dipole-allowed electronic transition spectrum is primarily photo-excited (``pumped'') by stellar Ly$\alpha$ photons~\citep{ardila02,herczeg02}. The Ly$\alpha$-pumping route proceeds primarily by absorption out of the second excited vibrational level ($v$~=~2) of H$_{2}$~\citep{shull78}, which implies that the molecules reside in a hot ($T(H_{2})$~$>$~2000 K) disk surface at semi-major axes $a$~$<$~10 AU~\citep{herczeg04}, or in extended outflows~\citep{walter03,saucedo03}. An analysis of the spectral line profiles can distinguish between these origins, enabling one to identify and characterize emission from the molecular disk given sufficient spectral resolution and spectroscopic sensitivity. Assuming that disk molecules are in Keplerian orbit around their central star, line broadening due to orbital motion dominates the profile in moderate-to-high inclination systems. H$_{2}$ velocity widths can therefore be used to infer the spatial distribution of the gas on the disk surface. 
\begin{figure}
\figurenum{1}
\begin{center}
\epsfig{figure=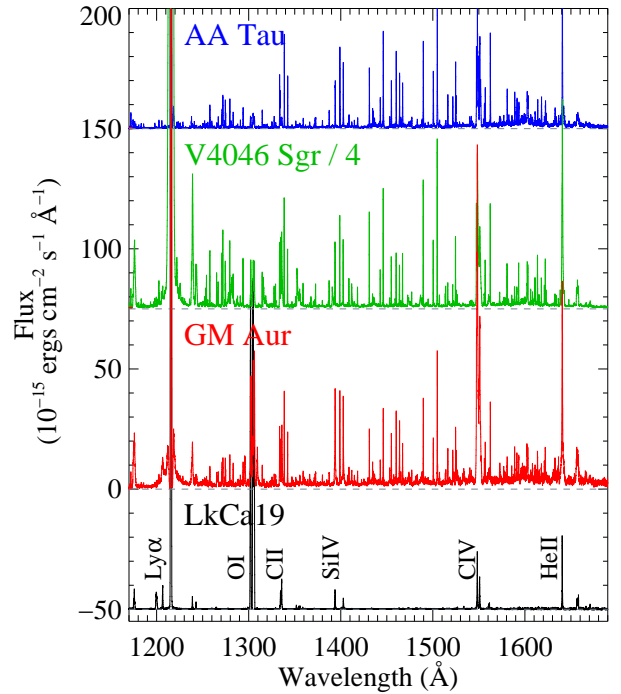,width=3.5in,angle=00}
\vspace{-0.1in}
\caption{
\label{cosovly} Examples of COS spectra (1170~--~1690~\AA) for a range of gas and dust disk parameters. From top to bottom: The primordial disk target AA Tau [offset by +150 ($\times$~10$^{-15}$ erg cm$^{-2}$ s$^{-1}$ \AA$^{-1}$)], the pre-transitional disk V4046 Sgr [flux divided by 4 and offset by +75 ($\times$~10$^{-15}$ erg cm$^{-2}$ s$^{-1}$ \AA$^{-1}$)], the transitional disk system GM Aur, and the gas-poor WTTS LkCa19 [offset by -50 ($\times$~10$^{-15}$ erg cm$^{-2}$ s$^{-1}$ \AA$^{-1}$)]. The spectra have been binned by one spectral resolution element (7 pixels) for display. Except for the atomic lines identified in the spectrum of LkCa19, most of the emission lines in the spectra of the other three stars are fluorescent H$_{2}$ emission lines pumped by Ly$\alpha$.}
\end{center}
\end{figure}

\begin{figure}
\figurenum{2}
\begin{center}
\epsfig{figure=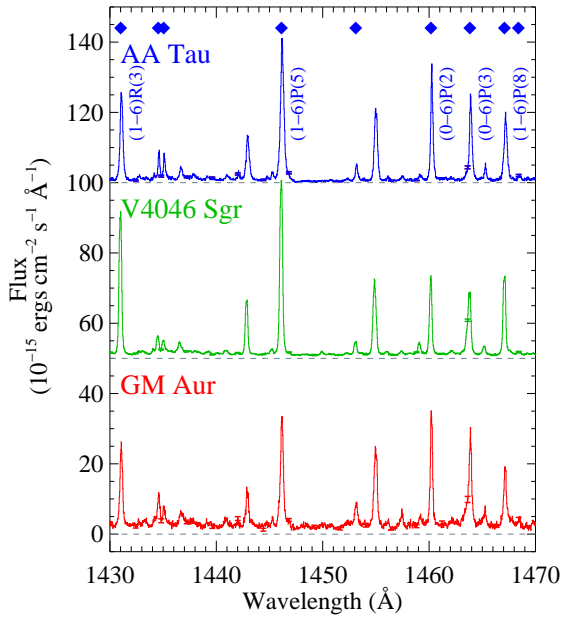,width=3.5in,angle=90}
\vspace{-0.1in}
\caption{The 1430~--~1470~\AA\ spectral region for the gas-rich targets plotted in Figure 1. All of the strong spectral features in this bandpass are emission lines from Ly$\alpha$-pumped fluorescent H$_{2}$. 
Emission lines used in this analysis are marked with blue diamonds and several bright features are labeled. 
Objects are plotted in order of decreasing near-IR dust excesses: AA Tau (primordial, $n_{13-31}$ = -0.51; Furlan et al. 2009), V4046 Sgr (pre-transitional) has a sub-AU scale hole in the inner disk dust distribution~\citep{jensen97}, and GM Aur (transitional, $n_{13-31}$ = 1.75; Furlan et al. 2009) has a $\sim$~24 AU hole in the inner dust disk~\citep{calvet05}. The spectra are binned to $\approx$~1/2 of a spectral resolution element (3 pixels), and representative error bars are shown overplotted. The H$_{2}$ spectra are qualitatively similar, independent of the inner disk dust properties. 
\label{cosovly}
}
\end{center}
\end{figure}

\begin{deluxetable*}{lccccccc}
\tabletypesize{\footnotesize}
\tablecaption{$HST$ Target List \label{lya_targets}}
\tablewidth{0pt}
\tablehead{ \colhead{Target\tablenotemark{a}} & \colhead{A$_V$} & \colhead{log$_{10}$(Age)} & \colhead{M$_*$} & \colhead{$\dot M$} & \colhead{$i$} & \colhead{HST\tablenotemark{a}} & \colhead{Ref.\tablenotemark{c}} \\ 
& & (yrs) & (M$_{\odot}$) & (10$^{-8}$ M$_{\odot}$ yr$^{-1}$) & ($\circ$) & PID & } 
\startdata
AA Tau & 0.50 & 6.38 $\pm$ 0.20 & 0.80 & 0.33 & 75 & 11616 & 2,15,28 \\ 
AK Sco & 0.5 & 7.24 $\pm$ 0.24 & 1.35 & 0.09 & 68 & 11616$-S$ & 3,24 \\ 
BP Tau & 0.50 & 5.94 $\pm$ 0.29 & 0.73 & 2.88 & 30 & 12036 & 1,15,30 \\ 
CS Cha & 0.8 & 6.39 $\pm$ 0.09 & 1.05 & 1.20 & 60 & 11616 & 4,25 \\ 
CV Cha & 1.67 & 6.70 $\pm$ 0.10 & 2.00 & 3.16 & 35 & 11616$-S$ & 5,26 \\ 
DE Tau & 0.60 & 5.82 $\pm$ 0.20 & 0.59 & 2.64 & 35 & 11616 & 1,15,29 \\ 
DF Tau A & 0.60 & 6.27 $\pm$ 0.53 & 0.19 & 17.7 & 85 & 11533 & 15,29 \\ 
DK Tau A & 0.80 & 6.17 $\pm$ 0.22 & 0.71 & 3.79 & 50 & 11616 & 1,15,29 \\ 
DM Tau & 0.0 & 6.56 $\pm$ 0.20 & 0.50 & 0.29 & 35 & 11616 & 2,16,21 \\ 
DN Tau & 1.90 & 6.04 $\pm$ 0.20 & 0.60 & 0.35 & 28 & 11616 & 2,15,21,31 \\ 
DR Tau & 3.20 & 6.18 $\pm$ 0.20 & 0.80 & 3.16 & 72 & 11616 & 2,17,28 \\ 
GM Aur & 0.10 & 6.86 $\pm$ 0.20 & 1.20 & 0.96 & 55 & 11616 & 2,15,21 \\ 
HD 104237 & 0.70 & 6.30 $\pm$ 0.30 & 2.50 & 3.50 & 18 & 11616$-S$ & 7,19,40 \\ 
HD 135344B & 0.30 & 6.90 $\pm$ 0.30 & 1.60 & 0.54 & 14 & 11828 & 8,19,36 \\ 
HN Tau A & 0.5 & 6.27 $\pm$ 0.27 & 0.85 & 0.13 & $>$40 & 11616 & 1,15,22 \\ 
IP Tau & 0.20 & 6.37 $\pm$ 0.24 & 0.68 & 0.08 & 60 & 11616 & 1,15,32 \\ 
LkCa 15 & 0.60 & 6.35 $\pm$ 0.26 & 0.85 & 0.13 & 49 & 11616 & 1,16,21 \\ 
RECX 11 & 0.0 & 6.60 $\pm$ 0.20 & 0.80 & 0.03 & 70 & 11616 & 9,20,45 \\ 
RECX 15 & 0.0 & 6.78 $\pm$ 0.08 & 0.40 & 0.10 & 60 & 11616 & 10,20 \\ 
RU Lupi & 0.07 & 6.39 $\pm$ 0.09 & 0.80 & 3.00 & 24 & 12036 & 11,18,34 \\ 
RW Aur A & 1.6 & 5.85 $\pm$ 0.53 & 1.40 & 3.16 & 77 & 11616 & 1,23,35 \\ 
SU Aur & 0.9 & 6.39 $\pm$ 0.21 & 2.30 & 0.45 & 62 & 11616 & 1,17,37 \\ 
SZ 102 & 1.13 & 6.15 $\pm$ 0.15 & 0.75 & 0.08 & 10 & 11616 & 12,27,38 \\ 
TW Hya & 0.0 & 7.00 $\pm$ 0.40 & 0.60 & 0.02 & 4 & 8041$-S$ & 13,18,36 \\ 
UX Tau A & 0.20 & 6.10 $\pm$ 0.30 & 1.30 & 1.00 & 35 & 11616 & 1,21 \\ 
V4046 Sgr & 0.0 & 6.90 $\pm$ 0.12 & 0.86+0.69 & 1.30 & 36 & 11533 & 14,22,39 \\ 
V836 Tau & 1.70 & 6.26 $\pm$ 0.26 & 0.75 & 0.01 & 65 & 11616 & 1,18,41 \\ 
\tableline
\tableline
HBC 427 & 0.00 & 6.64 $\pm$ 0.14 & 0.7 & $\cdots$ & $\cdots$ & 11616 & 6 \\ 
LkCa 19 & 0.00 & 6.84 $\pm$ 0.38 & 1.35 & $\cdots$ & $\cdots$ & 11616 & 1 \\ 
LkCa 4 & 0.69 & 6.43 $\pm$ 0.25 & 0.77 & $\cdots$ & $\cdots$ & 11616 & 6 \\ 
RECX 1 & 0.00 & 6.78 $\pm$ 0.00 & 0.90 & $\cdots$ & $\cdots$ & 11616 & 42 \\ 
TWA 13A & 0.00 & 6.90 $\pm$ 0.12 & 0.32 & $\cdots$ & $\cdots$ & 12361 & 43 \\ 
TWA 13B & 0.00 & 6.90 $\pm$ 0.12 & 0.38 & $\cdots$ & $\cdots$ & 12361 & 43 \\ 
TWA 7 & 0.00 & 6.39 $\pm$ 0.39 & 0.55 & $\cdots$ & $\cdots$ & 11616 & 44 
\enddata
\tablenotetext{a}{Targets in the upper group are CTTSs, targets in the lower group are WTTSs.} 
\tablenotetext{b}{Program IDs marked $-S$ indicate that STIS observations were used} 
\tablenotetext{c}{\rule{0mm}{5mm}
(1) \citet{Kraus2009}; (2) \citet{Ricci2010}, age uncertainties are assumed to be $\pm$~0.20; (3) \citet{Alencar2003}; (4) \citet{Lawson1996}; (5) \citet{Siess2000}; (6) \citet{Bertout2007}; (7) \citet{Feigelson2003}; (8) \citet{VanBoekel2005}; (9) \citet{Lawson2001}; (10) \citet{ramsay07}; (11) \citet{Herczeg2005}; (12) \citet{Comeron2010}; (13) \citet{Webb1999}; (14) \citet{Quast2000}; (15) \citet{Gullbring1998}; (16) \citet{Hartmann1998}; (17) \citet{Gullbring2000}; (18) \citet{Herczeg2008}; (19) \citet{Garcia2006}; (20) \citet{Lawson2004}; (21) \citet{Andrews2011}; (22) \citet{france11a}; (23) \citet{White2001}; (24) \citet{Gomez2009}; (25) \citet{Espaillat2007}; (26) \citet{Hussain2009}; (27) \citet{Comeron2003}; (28) \citet{Andrews2007}; (29) \citet{JKrull2001}; (30) \citet{Simon2000}; (31) \citet{Muzerolle2003}; (32) \citet{Espaillat2010}; (34) \citet{Stempels2007}; (35) \citet{Eisner2007}; (36) \citet{Pontoppidan2008}; (37) \citet{Akeson2002}; (38) \citet{Coffey2004}; (39) \citet{Rodriguez2010}; (40) \citet{Grady2004}; (41) \citet{Najita2008}; (42) \citet{ingleby11}; (43) \citet{Plavchan2009}; (44) \citet{Neuhauser2000}; (45) \citet{ingleby11b}}
\end{deluxetable*}

In this work, we present the most sensitive survey of spectrally resolved H$_{2}$ emission in protoplanetary disks obtained to date. Previous spectral surveys of TTSs with the {\it International Ultraviolet Explorer}~\citep{valenti00,krull00} and the various ultraviolet spectrographs on $HST$~\citep{ardila02,herczeg06,ingleby09,yang12} have been carried out at either lower spectroscopic sensitivity or resolution. In this study, we take advantage of the high sensitivity, low instrumental background, and moderate spectral resolution of the {\it Hubble Space Telescope}-Cosmic Origins Spectrograph to greatly expand the number of targets available for detailed UV studies. In \S2, we describe the targets and the $HST$ observations. The analysis performed to characterize the H$_{2}$ luminosities and spatial distributions are described in \S3. We present in \S4 a discussion of H$_{2}$ line profiles, considering outflow and disk origins for the emitting gas, and an estimate of the fraction of stellar Ly$\alpha$ re-processed by circumstellar H$_{2}$. \S4 also presents the time-evolution of the amount and location of the H$_{2}$ gas, suggesting that the H$_{2}$-emitting gas both dissipates and moves towards larger orbital radii over the interval from 10$^{6}$~--~10$^{7}$ yr.
A brief summary of the results from the molecular survey are presented in \S5. 


\section{Target Sample and Observations} 

\subsection{Target Sample} 

The goal of this observational survey is to span a range of ages, mass accretion rates, and star-forming environments in order to better understand the global properties of H$_{2}$ emission in protoplanetary environments. The targets mainly belong to the Taurus-Auriga, $\eta$ Chamaeleontis, TW Hya, and Chamaeleon I star-forming regions, as well as individual targets in other associations and isolated systems. Potential sources of uncertainty in analyzing a diverse population of targets are systematic effects based on different methods used to derive system parameters in the literature. In order to mitigate the effects of systematics on this study of molecular disks, our approach was to adopt system parameters from papers where similar techniques were used to derive properties such as ages, extinctions, stellar masses, and inclinations. The adopted target parameters are given in Table 1. 
Where possible, 1) inclinations were taken from sub-mm/IR interferometric studies, 2) ages, stellar masses, and extinctions were derived from pre-main-sequence stellar evolutionary tracks, and 3) mass accretion rates were derived from measurements of the accretion luminosity. Of our 34 targets, 27 are considered CTTSs while 7 do not show evidence for active accretion or a gas-rich circumstellar disk and are classified as Weak-lined T Tauri Star (WTTSs). 
These populations are separated as the upper and lower groups, respectively, of target stars listed in Table 1. 
7 of the CTTSs in our sample are considered transitional systems (CS Cha, DM Tau, GM Aur, UX Tau A, LkCa15, HD 135344B and TW Hya), with mid-IR SEDs indicating that a gap has opened in their inner dust disks. This sample includes several of the best-studied transitional disks in the literature.
References are listed in Table 1, and a more detailed description of a subsample of these targets can be found in~\citet{schindhelm11}. 

The assumed distances are not critical to the results presented here, but they 
do impact the comparison of the H$_{2}$, Ly$\alpha$, and \ion{C}{4} luminosities presented in \S4. 
For the Taurus-Auriga targets (AA Tau, BP Tau, DE Tau, DF Tau, DK Tau, DM Tau, DN Tau, DR Tau, GM Aur, HN Tau, IP Tau, LkCa15, RW Aur, SU Aur, UX Tau A, V836 Tau, HBC 427 (V397 Tau), LkCa19, LkCa4), we assumed $d$~=~140 pc~(Elias 1978; Kenyon \& Hartmann 1995; and see also the VLBA work presented by Loinard et al. 2007); for the $\eta$ Cha targets (RECX-11, RECX-15, RECX-1), we assumed $d$~=~97 pc~\citep{mamajek99}; for the TW Hya association targets (TW Hya, TWA13A, TWA13B, TWA7), we assumed $d$~=~54 pc~\citep{leeuwen07}, and for Chamaeleon I (CS Cha, CV Cha), we assumed $d$~=~160 pc~\citep{luhman04b}.  Other objects are V4046 Sgr ($d$~=~83 pc; Quast et al. 2000), SZ 102 (in Lupus 3, $d$~=~200 pc; Comer{\'o}n et al. 2003), HD104237 (a member of the $\epsilon$ Cha group, $d$~=~116 pc; Feigelson et al. 2003), AK Sco ($d$~=~103 pc; van Leeuwen 2007), and RU Lupi ($d$~=~121 pc; van Leeuwen et al. 2007). Six of the nineteen sources in Taurus-Auriga are known multi-star systems (see, e.g., Kraus et al. 2012) and V4046 Sgr is known to be a short-period binary system~\citep{quast00}. 
\nocite{quast00,comeron03,feigelson03,leeuwen07,herczeg05,elias78,kenyon95,loinard07,kraus12}

\begin{figure*}
\figurenum{3a}
\begin{center}
\epsfig{figure=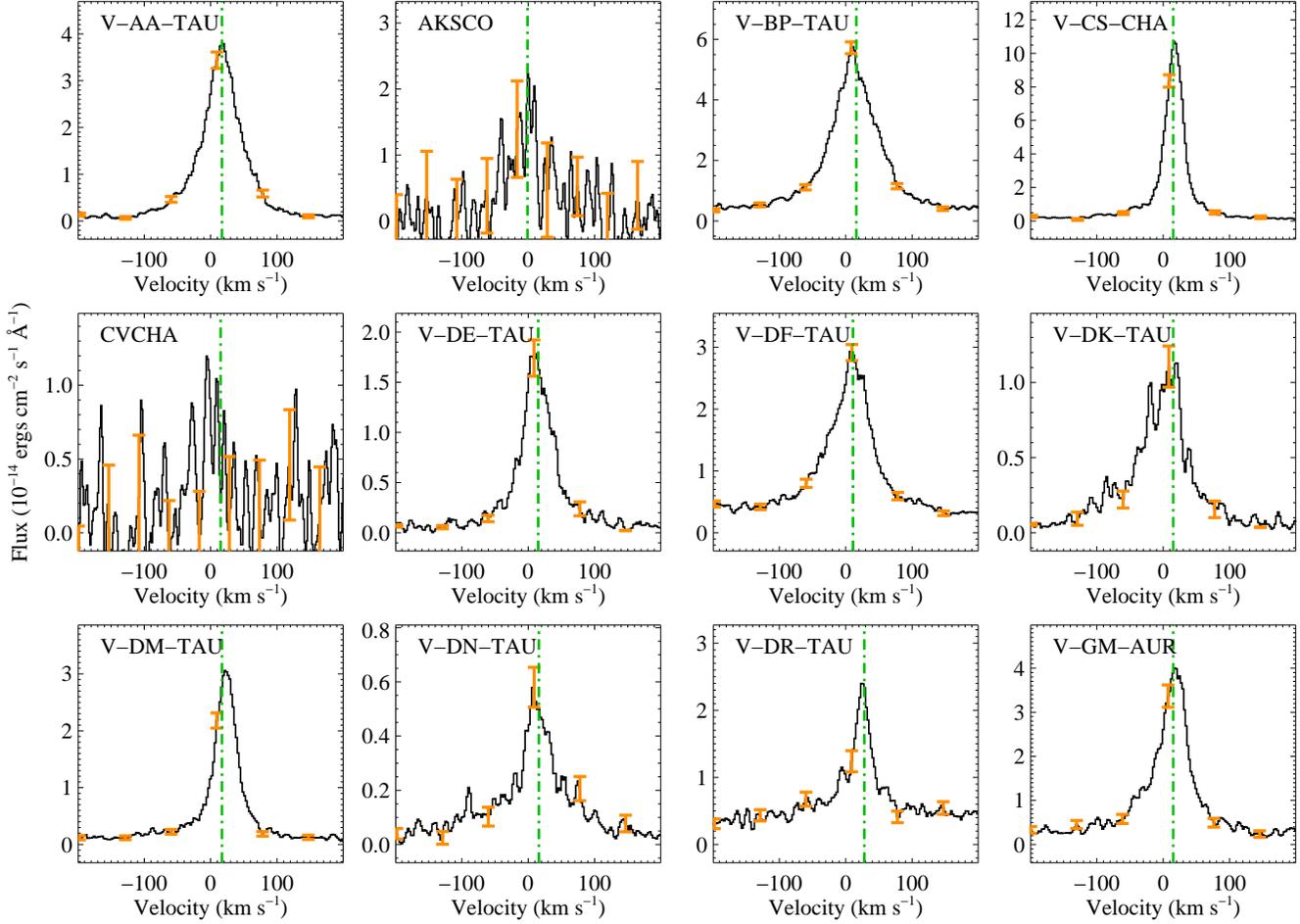,width=5.0in,angle=90}
\vspace{+0.1in}
\caption{
\label{cosovly} Velocity profiles for the H$_{2}$ $B$$^{1}\Sigma^{+}_{u}$~--~$X$$^{1}\Sigma^{+}_{g}$ (1~--~7) R(3) ($\lambda_{lab}$~=~1489.57~\AA) emission line in all targets, plotted over the $\pm$ 200 km s$^{-1}$ interval. The data plotted here are smoothed to 3 pixels ($\approx$~0.5 spectral resolution elements), with representative error bars shown in orange. The green dash-dotted line is plotted at the stellar radial velocity, when known. 
The target labels are the target names identified in the headers of the $HST$ spectroscopic observations. 
}
\end{center}
\end{figure*}

\begin{figure*}
\figurenum{3b}
\begin{center}
\epsfig{figure=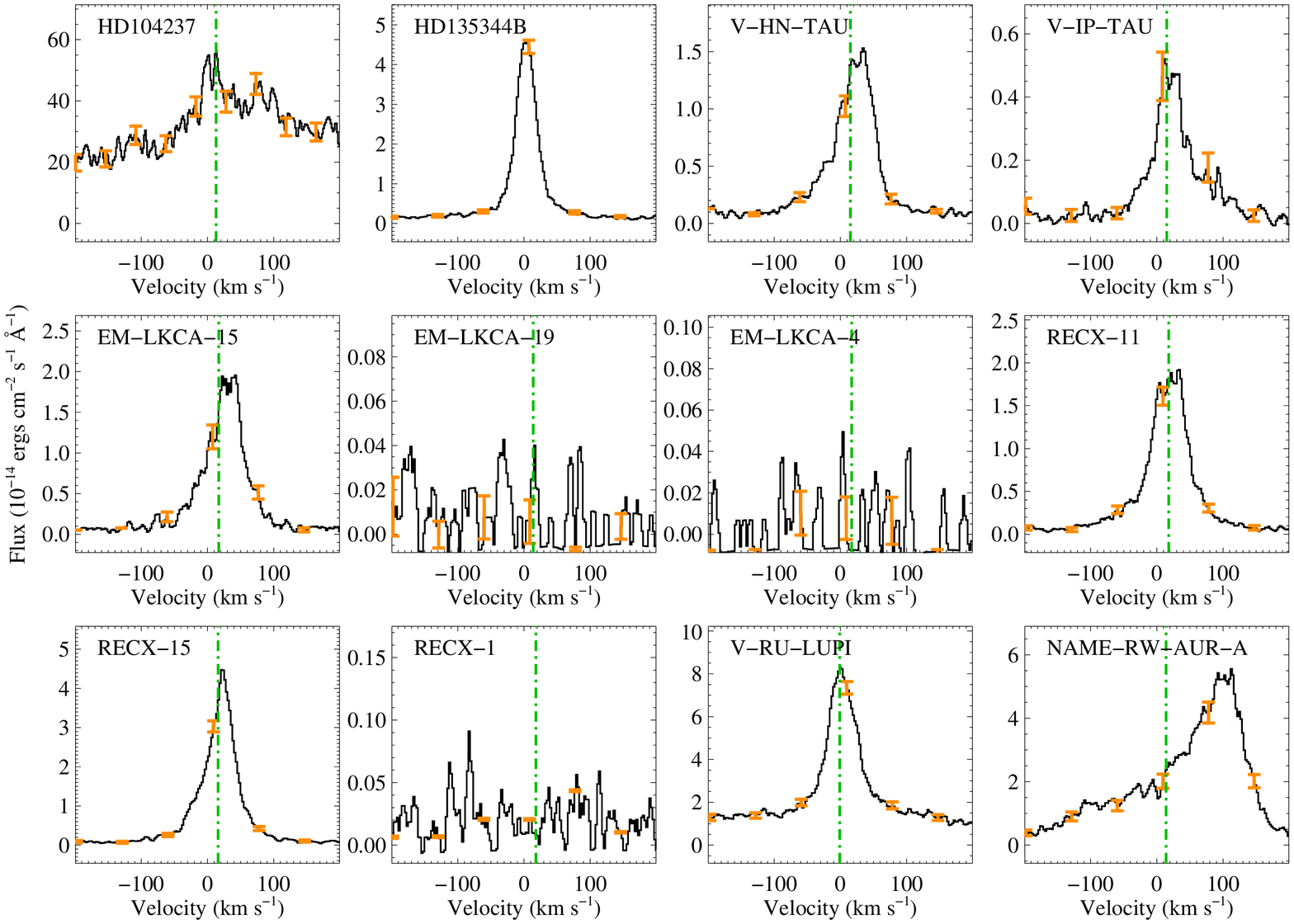,width=5.0in,angle=90}
\vspace{+0.1in}
\caption{
\label{cosovly} same as Figure 3a.
}
\end{center}
\end{figure*}

\begin{figure*}
\figurenum{3c}
\begin{center}
\epsfig{figure=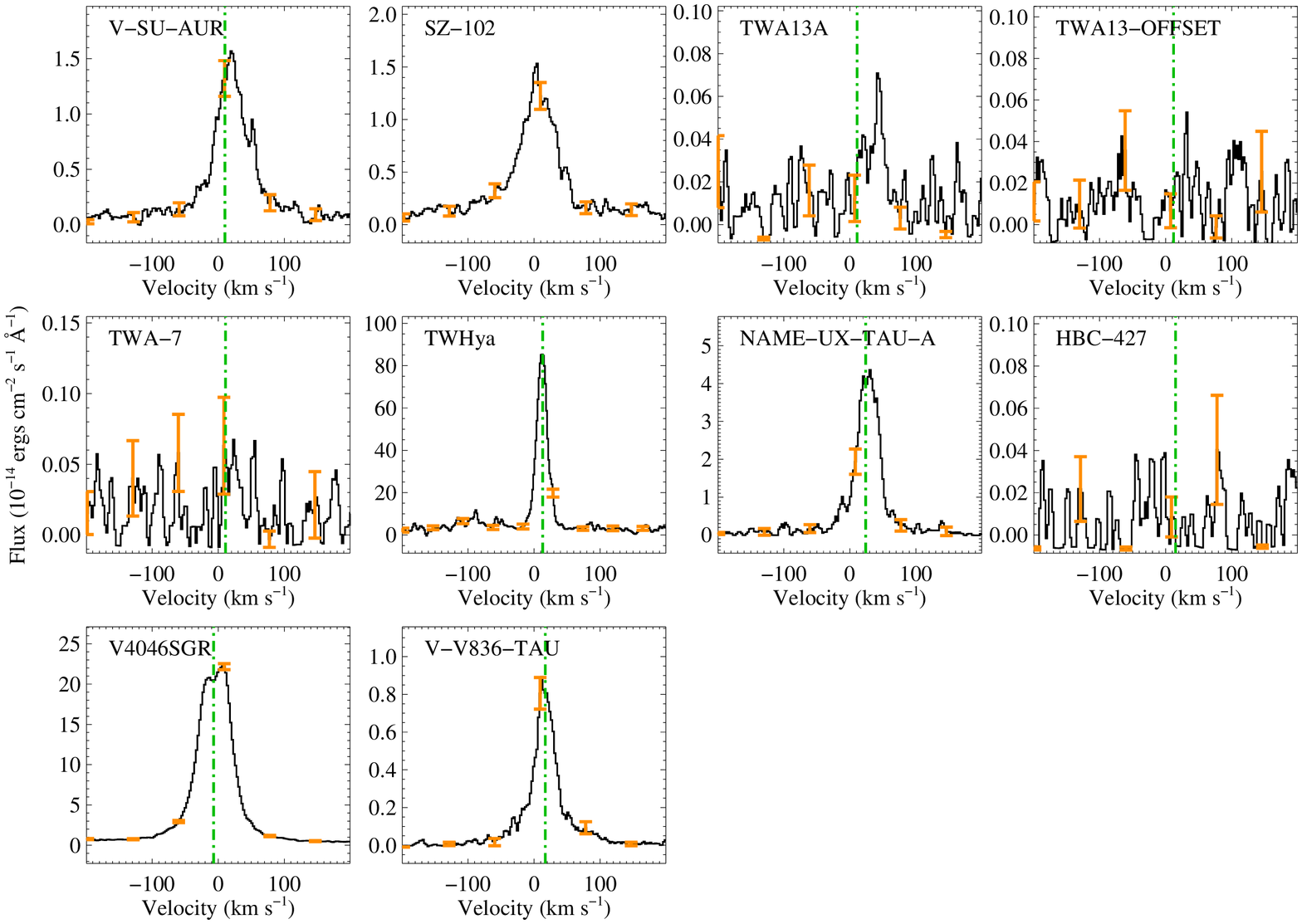,width=5.0in,angle=90}
\vspace{+0.1in}
\caption{
\label{cosovly} same as Figure 3a.
}
\end{center}
\end{figure*}

\begin{deluxetable}{lcccc}
\tabletypesize{\scriptsize}
\tablecaption{Selected H$_{2}$ Emission Lines. \label{lya_lines}}
\tablewidth{0pt}
\tablehead{
\colhead{Line ID\tablenotemark{a}} & \colhead{[$v^{'}$,$J^{'}$]} & \colhead{$\lambda_{lab}$ (\AA)} & \colhead{$B_{mn}$\tablenotemark{b}} & \colhead{$\lambda_{pump}$ (\AA)} }
\startdata
(3~--~9)P(14) & [3,13] & 1608.33 & 0.139 & 1213.36 \\
(3~--~10)R(12) & [3,13] & 1615.43 & 0.125 & 1213.36 \\
\tableline
(4~--~6)R(12) & [4,13] & 1415.33 & 0.037 & 1213.68 \\
(4~--~8)R(12) & [4,13] & 1509.45 & 0.023 & 1213.68 \\
(4~--~11)R(12) & [4,13] & 1613.99 & 0.092 & 1213.68 \\
\tableline
(3~--~5)R(15) & [3,16] & 1418.23 & 0.050 & 1214.47 \\
(3~--~7)R(15) & [3,16] & 1513.99 & 0.057 & 1214.47 \\
(3~--~9)R(15) & [3,16] & 1593.26 & 0.122 & 1214.47 \\
(3~--~10)R(15) & [3,16] & 1621.12 & 0.062 & 1214.47 \\
\tableline
(4~--~8)P(5) & [4,4] & 1477.05 & 0.039 & 1214.78 \\
(4~--~9)P(5) & [4,4] & 1526.55 & 0.033 & 1214.78 \\
(4~--~11)P(5) & [4,4] & 1613.72 & 0.150 & 1214.78 \\
\tableline
(1~--~6)P(8) & [1,7] & 1467.08 & 0.080 & 1215.73 \\
(1~--~7)R(6) & [1,7] & 1500.45 & 0.101 & 1215.73 \\
(1~--~7)P(8) & [1,7] & 1524.65 & 0.111 & 1215.73 \\
(1~--~8)R(6) & [1,7] & 1556.87 & 0.074 & 1215.73 \\
\tableline
(1~--~6)R(3) & [1,4] & 1431.01 & 0.058 & 1216.07 \\
(1~--~6)P(5) & [1,4] & 1446.12 & 0.083 & 1216.07 \\
(1~--~7)R(3) & [1,4] & 1489.57 & 0.094 & 1216.07 \\
(1~--~7)P(5) & [1,4] & 1504.76 & 0.115 & 1216.07 \\
\tableline
(3~--~7)P(1) & [3,0] & 1435.05 & 0.118 & 1217.04 \\
(3~--~10)P(1) & [3,0] & 1591.32 & 0.233 & 1217.04 \\
(3~--~11)P(1) & [3,0] & 1636.34 & 0.099 & 1217.04 \\
\tableline
(0~--~5)P(2) & [0,1] & 1398.95 & 0.141 & 1217.21 \\
(0~--~6)P(2) & [0,1] & 1460.17 & 0.083 & 1217.21 \\
(0~--~2)P(2) & [0,1] & 1521.59 & 0.032 & 1217.21 \\
\tableline
(0~--~5)P(3) & [0,2] & 1402.65\tablenotemark{c} & 0.126 & 1217.64 \\
(0~--~6)P(3) & [0,2] & 1463.83 & 0.074 & 1217.64 \\
(0~--~7)P(3) & [0,2] & 1525.15 & 0.029 & 1217.64 \\
\tableline
(2~--~5)P(13) & [2,12] & 1434.54 & 0.066 & 1217.90 \\
(2~--~6)R(11) & [2,12] & 1453.10 & 0.049 & 1217.90 \\
(2~--8)R(11) & [2,12] & 1555.89 & 0.077 & 1217.90 \\
(2~--~8)P(13) & [2,12] & 1588.80 & 0.119 & 1217.90 \\
\tableline
(2~--~8)P(16) & [2,15] & 1612.39 & 0.138 & 1218.52 \\
(2~--~9)R(14) & [2,15] & 1617.42 & 0.103 & 1218.52 \\
\tableline
(0~--~5)R(2) & [0,3] & 1395.20 & 0.096 & 1219.09 \\
(0~--~5)P(4) & [0,3] & 1407.29 & 0.120 & 1219.09 \\
(0~--~6)P(4) & [0,3] & 1468.39 & 0.070 & 1219.09
\enddata
\tablenotetext{a}{Transitions are for the $B$$^{1}\Sigma^{+}_{u}$~--~$X$$^{1}\Sigma^{+}_{g}$ H$_{2}$ band system. } 
\tablenotetext{b}{The branching ratio is the ratio of the line transition probability to the total transition probability out of state [$v^{'}$,$J^{'}$], $B_{mn}$ = $\frac{A_{n'v'J' \rightarrow v''J''}}{A_{n'v'J'}}$ } 
\tablenotetext{c}{Blended with \ion{Si}{4} $\lambda_{lab}$~=~1402.77~\AA\ in some targets. } 
\end{deluxetable}

\subsection{Observations}

Our sample of 34 T Tauri stars was assembled from new and archival observations with $HST$-COS and -STIS. The majority of the targets were observed as part of the DAO of Tau guest observing program (PID 11616; PI - G. Herczeg) and the COS Guaranteed Time Observing program (PIDs 11533 and 12036; PI - J. Green). Additional observations of the transitional disk HD135344B and weak-lined systems TWA13A and TWA13B (PIDs 11828 and 12361; PI - A. Brown) are presented. A subset of the H$_{2}$ survey observations have been presented previously in the literature~\citep{france11a,france11b,france12a,ingleby11,yang11,schindhelm11}. Finally, we have included archival STIS observations of the well-studied CTTS TW Hya~\citep{herczeg02}, obtained through StarCAT~\citep{ayres10}. 

\begin{figure}
\figurenum{4}
\begin{center}
\epsfig{figure=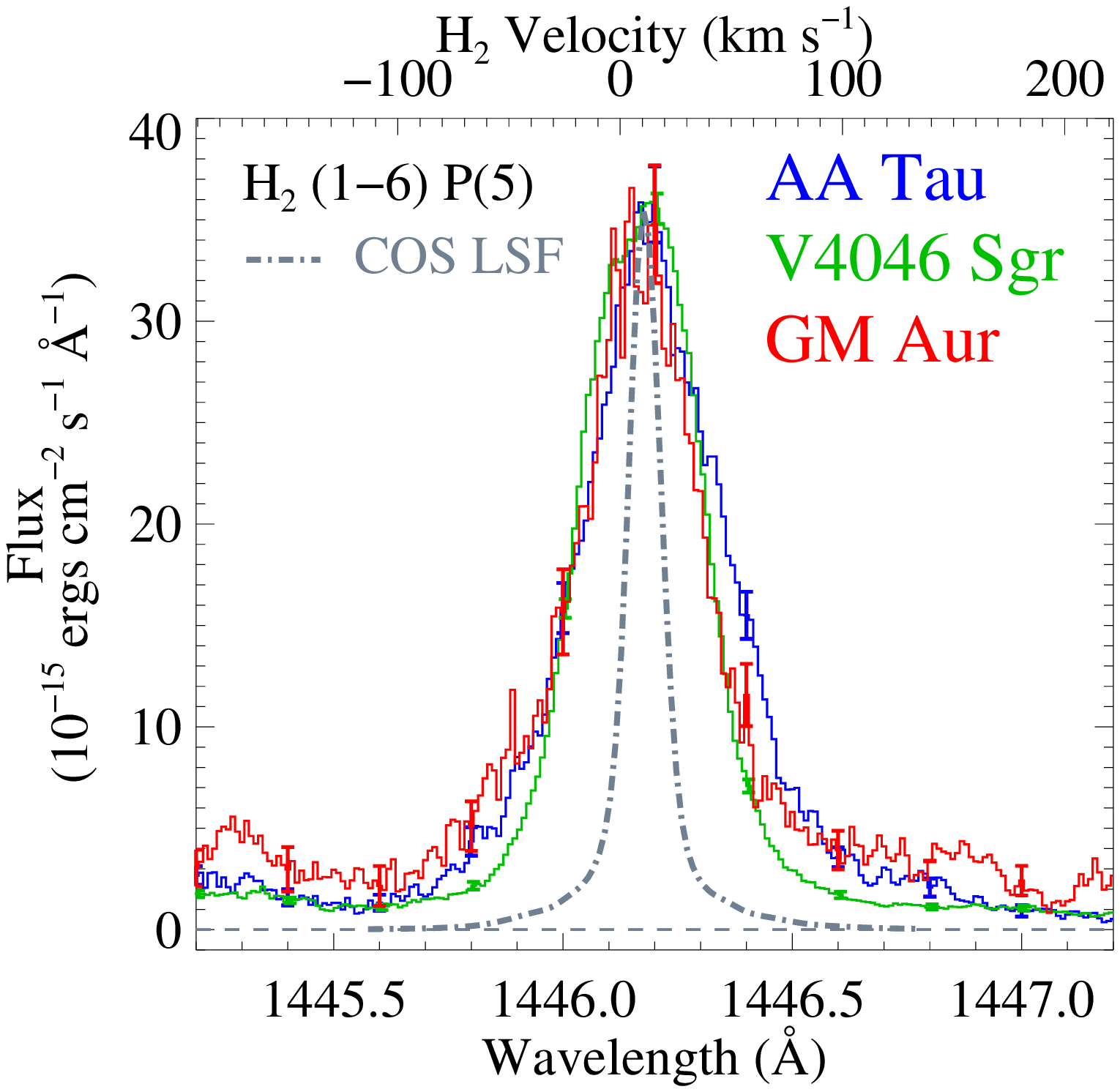,width=3.0in,angle=90}
\vspace{-0.2in}
\caption{
\label{cosovly} Expanded view of the H$_{2}$ (1~--~6) P(5) ($\lambda_{lab}$~=~1446.12\AA) emission line, showing the velocity width beyond the COS linespread function (shown as the gray dash-dotted line) for the three example targets shown in Figure 2. The AA Tau and V4046 Sgr spectra have been normalized to the peak flux of the GM Aur H$_{2}$ emission and shifted to the centroid velocity of the H$_{2}$ emission. The emission lines are spectrally resolved by COS.
}
\end{center}
\end{figure}

Most of the targets were observed with the medium-resolution far-UV modes of COS (G130M and G160M; Green et al. 2012).~\nocite{green12} These observations were acquired between 2009 December and 2011 September. Multiple central wavelength settings at several focal-plane split positions were used to create continuous far-UV spectra from $\approx$ 1150~--~1750~\AA\ and mitigate the effects of fixed pattern noise. These modes provide a point-source resolution of $\Delta$$v$~$\approx$~17 km s$^{-1}$ with 7 pixels per resolution element~\citep{osterman11}. The data were smoothed by three pixels for analysis. The total far-UV exposure times were between two and four orbits per target, depending on the intrinsic luminosity and the interstellar plus circumstellar reddening on the sightline. The one-dimensional spectra produced by the COS calibration pipeline, CALCOS, were aligned and coadded using the custom software procedure described by~\citet{danforth10}. The full far-UV spectra of three CTTSs (AA Tau, V4046 Sgr, and GM Aur) and one WTTS (LkCa19) are displayed in Figure 1, and a 40~\AA\ blow-up of the CTTSs is shown in Figure 2.

Targets that exceeded the COS bright-object limit were observed with STIS in the medium-resolution echelle mode. We used the E140M mode ($\Delta$$v$~$\approx$~8 km s$^{-1}$) mode through the 0.2\arcsec~$\times$~0.2\arcsec\ slit for exposure times between two and three orbits per object. The far-UV STIS spectra were combined using the STIS echelle software developed for the StarCAT catalog (T. Ayres~--~private communication; Ayres 2010). 

\begin{figure*}
\figurenum{5}
\begin{center}
\epsfig{figure=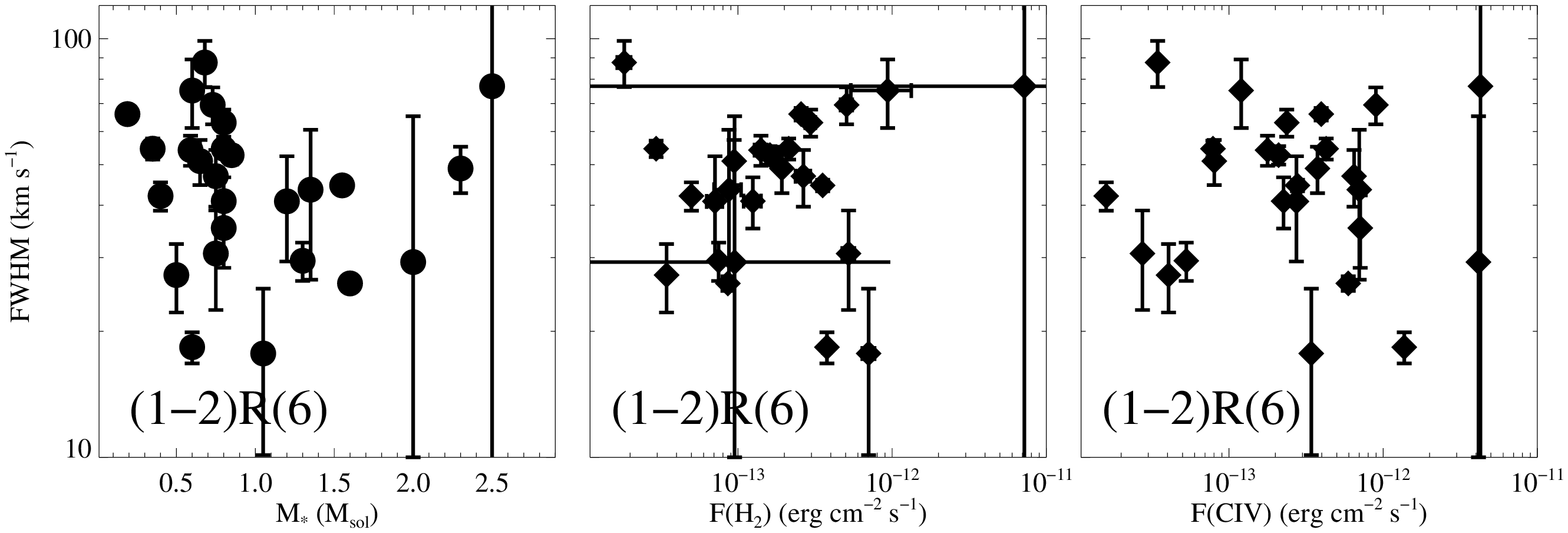,width=2.5in,angle=90}
\epsfig{figure=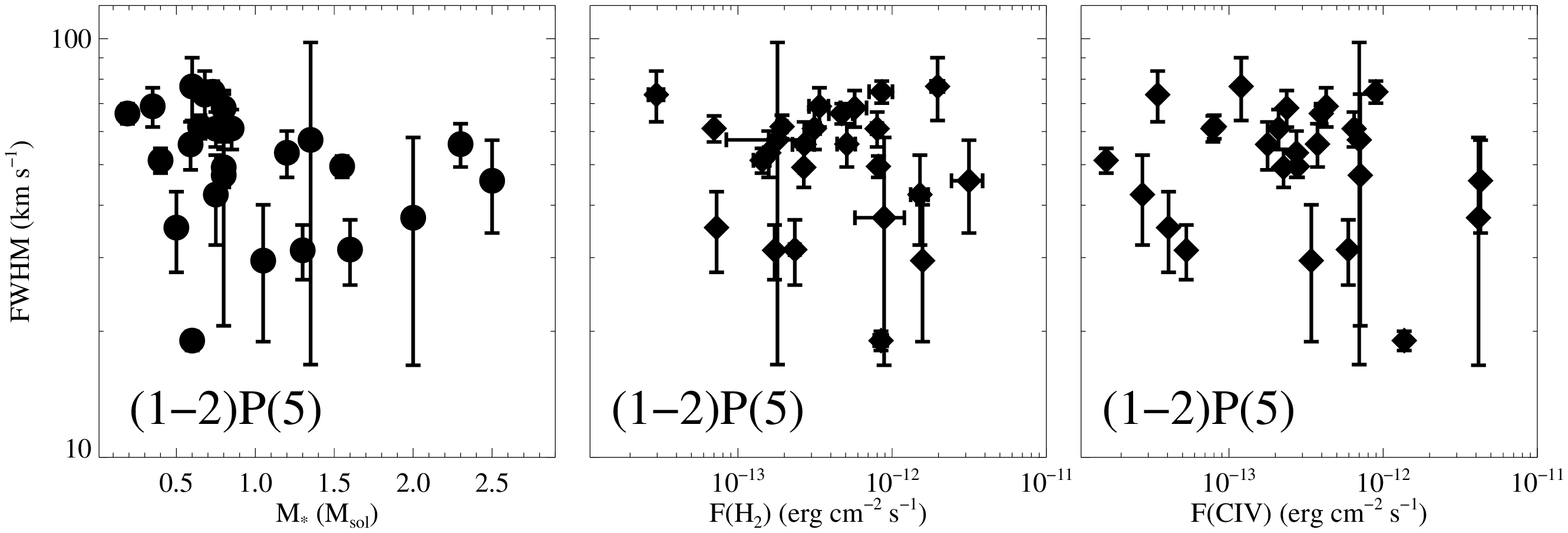,width=2.5in,angle=90}
\caption{
\label{cosovly} The average H$_{2}$ FWHM is compared with the stellar mass, the progression flux ($F_{m}$(H$_{2}$)), and the \ion{C}{4} flux ($F$(\ion{C}{4})) for two progressions: [1,7] (pumped through the (1~--~2) R(6) $\lambda_{lab}$~=~1215.73\AA\ transition) and [1,4] (pumped through the (1~--~2) P(5) $\lambda_{lab}$~=~1216.07\AA\ transition). The H$_{2}$ emission line FWHMs are uncorrelated with the stellar mass, $F_{m}$(H$_{2}$), or $F$(\ion{C}{4}). 
}
\end{center}
\end{figure*}


\section{Analysis}

We observe fluorescent H$_{2}$ emission from all of the 27 CTTSs in our sample, and no H$_{2}$ emission from the 7 WTTS targets. The number of observed fluorescent progressions varies significantly across the sample, and we present our measurements of the total H$_{2}$ fluxes below. We detect strong fluorescent emission in all of the transitional objects in our sample (CS Cha, DM Tau, GM Aur, UX Tau A, LkCa15, HD 135344B and TW Hya; \S4.3). 

The fluorescent H$_{2}$ lines observed in the CTTS sample can be used to determine the relative amount of H$_{2}$ in the circumstellar environment and to constrain its spatial distribution. 
For our line-profile analysis, we focus on the measurement of two progressions ([$v^{'}$,$J^{'}$] = [1,7] and [1,4]) detected in all of the CTTS targets\footnote{The quantum numbers $v$ and $J$ denote the vibrational and rotational quantum numbers in the ground electronic state ($X$$^{1}\Sigma^{+}_{g}$), the numbers $v^{'}$ and $J^{'}$ characterize the H$_{2}$ in the excited ($B$$^{1}\Sigma^{+}_{u}$) electronic state, and the numbers $v^{''}$ and $J^{''}$ are the rovibrational levels in the electronic ground state following the fluorescent emission. Absorption lines are described by ($v^{'}$~--~$v$) and emission lines by ($v^{'}$~--~$v^{''}$).}. These emission lines are pumped through the (1~--~2)R(6) $\lambda_{lab}$1215.73~\AA\ and (1~--~2)P(5) $\lambda_{lab}$1216.07~\AA\ transitions, respectively. The absorbing transitions are within +15~--~+100 km s$^{-1}$ of Ly$\alpha$ line-center. 
The signal-to-noise ratios (S/N) per resolution element are typically between 5 and 40 in the brightest fluorescent H$_{2}$ emission lines for our CTTS targets. The (1~--~7)R(3) $\lambda_{lab}$~=~1489.57~\AA\ transition is relatively free from spectral contamination and is displayed for all targets in Figure 3 ($a$~--~$c$). 
When available, the stellar radial velocities are indicated in Figure 3 with green dash-dotted lines. Within the $\sim$~15 km s$^{-1}$ wavelength solution accuracy of COS, most of the H$_{2}$ progressions are consistent with the stellar velocity. 
In several cases 
the H$_{2}$ lines appear to have line wings extending to negative velocities or statistically significant differences in the velocity widths of different H$_{2}$ progressions. RW Aur is an extreme example of this behavior. \citet{ardila02} and \citet{herczeg06} have noted the presence of blue-shifted H$_{2}$ emission in their sample of T Tauri stars observed with GHRS and STIS. The presence of blue-shifted emission creates additional uncertainty in the measured line-widths at the resolution of our spectra, and we present a discussion of outflow signatures observed in our sample in \S4.2.1. 

The high S/N of the COS data means that we can restrict the analysis to the brightest lines from the progressions of interest. Systematics were minimized by focusing on emission lines with wavelengths 1395~$\lesssim$~$\lambda$~$\lesssim$~1640~\AA. This choice is optimal because 1) fluorescent transitions cascading to vibrational levels $v^{''}$~$\gtrsim$~5 do not suffer significant self-absorption before escaping the circumstellar environment (e.g., Figure 7 of Herczeg et al. 2004), enabling more robust flux measurements, 
2) the non-Gaussian wings of the COS instrumental line-spread function (LSF\footnote{The LSF experiences a wavelength dependent non-Gaussianity due to the introduction of mid-frequency wave-front errors produced by the polishing errors on the $HST$ primary and secondary mirrors; {\tt http://www.stsci.edu/hst/cos/documents/isrs/}}) contain a smaller fraction of the power at $\lambda$~$>$~1400~\AA, 3) the strongest lines from the [1,4] and [1,7] progressions are in this range, and 
4) the relative correction for interstellar reddening between the different fluorescent emission lines is minimized. The first and latter two arguments apply equally to the STIS data. 

\begin{figure*}
\figurenum{6}
\begin{center}
\epsfig{figure=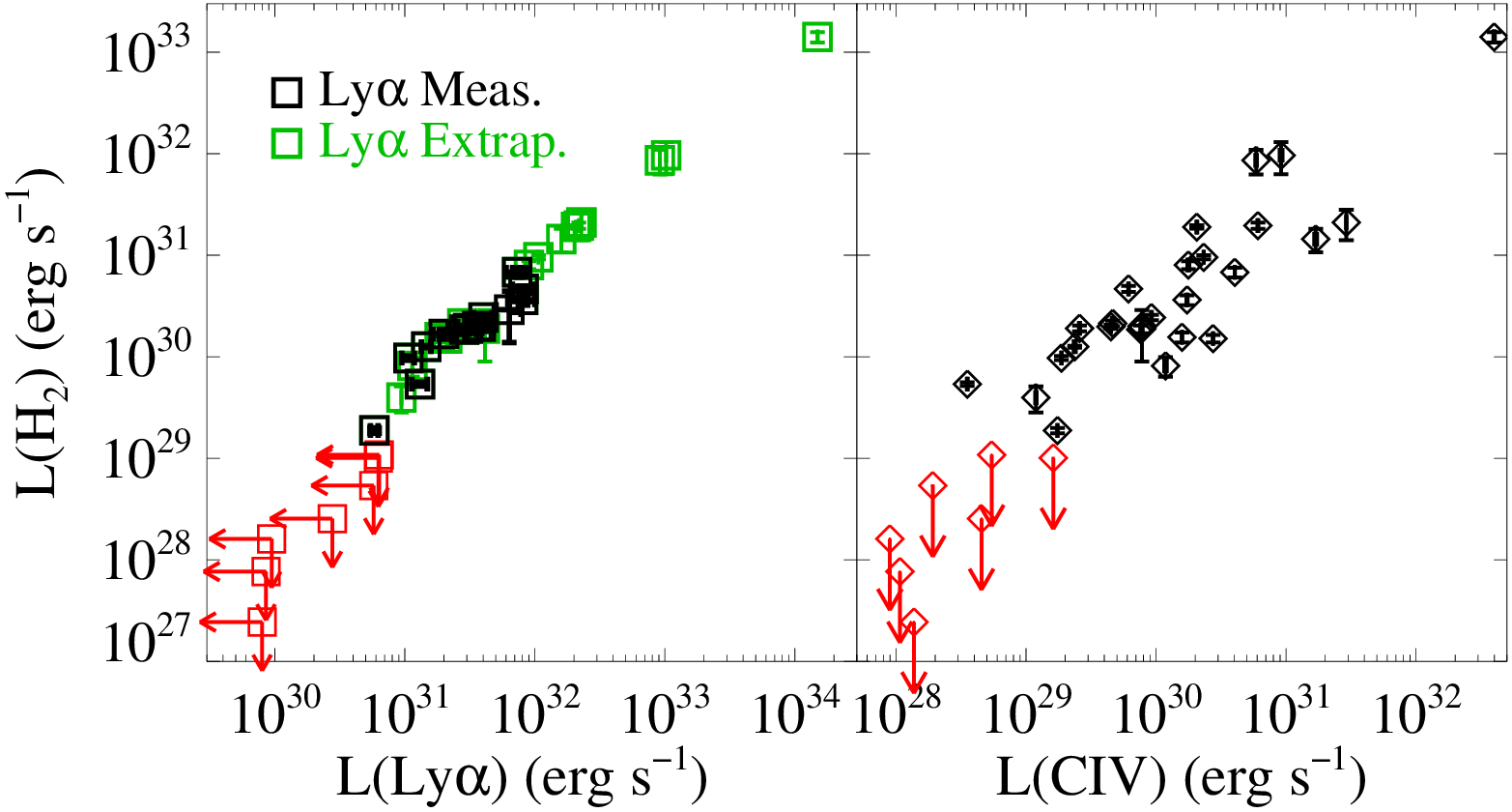,width=4.0in,angle=90}
\vspace{-0.1in}
\caption{
\label{cosovly} A comparison of the hot gas [log$_{10}$($T_{form}$)~$\sim$~4~--~5] and molecular gas emission from the TTSs studied here. The Ly$\alpha$ and \ion{C}{4} emission is produced in the protostellar atmosphere through a combination of magnetic activity and magnetically funneled accretion. The H$_{2}$ resides in the circumstellar disk and in some cases, extended outflows, excited by Ly$\alpha$ photons. The H$_{2}$ luminosity is the summed luminosity of the 12 progressions listed in Table 2, representing $\gtrsim$~80\% of the total H$_{2}$ luminosity (depending on the specific Ly$\alpha$ pumping profile). Targets with H$_{2}$ upper limits are plotted in red. The black squares in the left plot are our H$_{2}$ measurements and the empirically determined Ly$\alpha$ fluxes presented by~\citet{schindhelm12a}. The green squares are an extrapolation of the $F$(H$_{2}$) vs. $F$(Ly$\alpha$) relationship. 
}
\end{center}
\end{figure*}

\begin{figure}
\figurenum{7}
\begin{center}
\epsfig{figure=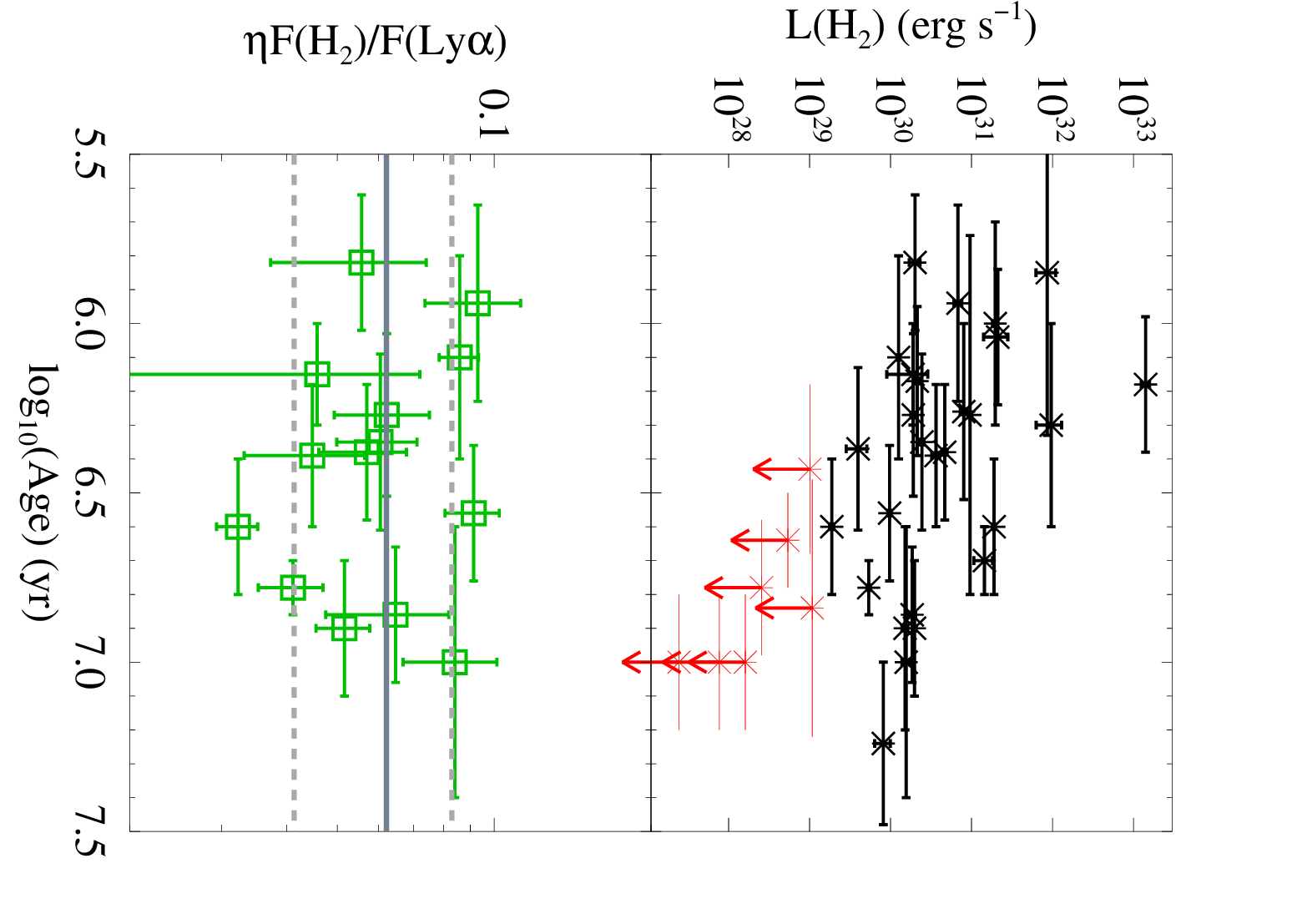,width=5.5in,angle=90}
\vspace{-0.1in}
\caption{
\label{cosovly} We plot the time evolution of the H$_{2}$ luminosity from the protoplanetary environment ($top$). The H$_{2}$ detections are shown in black and the non-detections in red. The lower plot 
presents a measure of the H$_{2}$ reprocessing of the stellar Ly$\alpha$ emission, expressed as the H$_{2}$ flux, divided by the empirically determined photo-exciting Ly$\alpha$ fluxes~\citep{schindhelm12a}. $\eta$ is a correction for radiative transfer effects and the anisotropy of the system, set here to $\eta$~=~1 (see \S4.1). The average H$_{2}$ reprocessing fraction of the total Ly$\alpha$ profile is 6.2~$\pm$~2.1~\%, assuming $\eta$~=~1, shown with solid (mean) and dashed (1~$\sigma$) gray lines in the lower panel.
}
\end{center}
\end{figure}

\subsection{Line Fluxes and the H$_{2}$ Progression Luminosity}

The H$_{2}$ emission lines were fit with an interactive multi-Gaussian IDL line-fitting code optimized for COS emission line spectra. This code assumes a Gaussian line-shape convolved with the wavelength dependent LSF, then uses the MPFIT routine to minimize $\chi^{2}$ between the fit and data~\citep{markwardt09}. A second order polynomial background, the Gaussian amplitudes, and the Gaussian FWHMs for each component are free parameters. The parameters of the underlying Gaussian emission lines are returned to the user. The smaller STIS aperture does not sample the broad wings of the $HST$ LSF, therefore unconvolved Gaussians were used for the targets observed with STIS. 
We chose 12 progressions with absorbing transitions that spanned the width of the observed Ly$\alpha$ profiles in most targets, 1213.3~$\leq$~$\lambda_{abs}$(H$_{2}$) ~$\leq$~1219.1~\AA~\citep{schindhelm12a}. We fit the brightest unblended lines from 12 progressions in the 1395~--~1640~\AA\ bandpass, and used these values to determine the total flux from each progression. 

A list of the selected lines is given in Table 2. The total flux from a given progression is given by 
\begin{equation}
F_{m}(\textup{H$_{2}$}) = \frac{1}{N} \sum \left ( \frac{F_{mn}}{B_{mn}} \right ) 
\end{equation}
where $F_{mn}$ is the reddening corrected, integrated H$_{2}$ emission line flux 
from rovibrational state $m$ (= [$v^{'}$,$J^{'}$]) in the $B$$^{1}\Sigma^{+}_{u}$ electronic state to $n$ (= [$v^{''}$,$J^{''}$]) in the ground electronic state, $X$$^{1}\Sigma^{+}_{g}$. $B_{mn}$ is the branching ratio between these two states, and $N$ is the number of emission lines measured from a given progression. 
The measurement errors are typically small, so we take the flux error to be the standard deviation of the individual measurements of $F_{m}$(H$_{2}$).  The dominant systematic error on the measured H$_{2}$ flux is the correction for interstellar reddening; we do not attempt to account for this uncertainty in the flux and luminosity errors presented below.  See the ApJ version of this paper (or contact the authors) for reddening-corrected progression fluxes for all of the gas-rich targets. 
The total progression luminosity is then $L_{m}$(H$_{2}$)~=~(4$\pi$$d^{2}$)$F_{m}$(H$_{2}$). Upper limits on the H$_{2}$ emission line fluxes in the gas-depleted targets were determined from the standard deviation in a $\pm$~50 km s$^{-1}$ region surrounding the laboratory wavelength of the transition. The total fluorescent H$_{2}$ luminosity is then taken as the sum of all 12 of the fluorescent progressions. This prescription should account for $\gtrsim$~80~\% of the total Ly$\alpha$-pumped H$_{2}$ emission from our targets, although the exact fraction of the total measured H$_{2}$ flux will depend on the local Ly$\alpha$ line-profile (Herczeg et al. 2004; Schindhelm et al. 2012b).~\nocite{schindhelm12a} 

Depending on the geometry and spatial distribution of the absorbing molecular layer, the H$_{2}$ absorption lines may be optically thin (or have optical depths of a few) in some targets. In this case, the emitted H$_{2}$ luminosity will be directly proportional to the number of Ly$\alpha$ pumping photons received. 
~\citet{schindhelm12a} have presented total incident Ly$\alpha$ fluxes for 14 of the targets in our sample. Using a line-profile reconstruction technique that takes into account the 12 H$_{2}$ progressions described above, 
they simultaneously fit the neutral hydrogen outflow, the H$_{2}$ column density, and the H$_{2}$ temperature (previous examples of H$_{2}$-based Ly$\alpha$ reconstructions are described by Wood et al. 2002; Wood \& Karvoska 2004; Herczeg et al. 2004). 
The amount of the stellar Ly$\alpha$ that is redistributed by H$_{2}$ can be determined by dividing the total H$_{2}$ flux, $F$(H$_{2}$), by the total Ly$\alpha$ flux, $F$(H$_{2}$)/$F$(Ly$\alpha$). The H$_{2}$ luminosity can also be compared to accretion indicators in our data set, such as \ion{C}{4}. Emission from the 
\ion{C}{4} $\lambda\lambda$1548,1550~\AA\ resonance doublet in excess of a baseline magnetospheric level has been shown to correlate with the mass accretion rate~\citep{krull00,yang12}. The \ion{C}{4} flux can be measured directly in the same data set from which we measure $F_{m}$(H$_{2}$), mitigating complications associated with short- and long-baseline time variability. Due to the non-Gaussian appearance of many of the \ion{C}{4} emission profiles, we measured $F$(\ion{C}{4}) by integrating the reddening corrected spectra over (1547.5~--~1553.5~\AA) and subtracted the continuum from an adjacent, line-free portion of the spectrum. 

\subsection{H$_{2}$ Line Widths and the Average H$_{2}$ Radius, $\langle$$R_{H2}$$\rangle$}

Figure 4 shows the H$_{2}$ (1~--~6)P(5) ($\lambda_{lab}$~=~1446.12~\AA) profiles of the three example spectra displayed in Figure 2, with the 1450~\AA\ $HST$+COS LSF overplotted as the gray dash-dot line. These spectra are typical of the CTTS sample and one observes that the emission lines are spectrally resolved. We display the basic H$_{2}$ and \ion{C}{4} line parameter observations for the [1,7] and [1,4] progressions in Figure 5. The H$_{2}$ emission line FWHMs are uncorrelated with the stellar mass, $F_{m}$(H$_{2}$), or $F$(\ion{C}{4}). 

Kinematic broadening dominates the observed H$_{2}$ line profiles (see \S4.2). The thermal broadening of the emission lines is approximately 4.5 km s$^{-1}$ at the nominal 2500 K H$_{2}$ layer, significant additional broadening would require temperatures in excess of the $\approx$~4500 K dissociation temperature of H$_{2}$~\citep{lepp83}. If we further assume that any turbulence in the disks is subsonic, then the maximum turbulent velocity will be no larger than a few km s$^{-1}$. Therefore, velocity broadening due to bulk motions and Keplerian rotation dominate the observed line shapes when the FWHM of the emission line is greater than the 17 km s$^{-1}$ spectral resolution of COS. 
For the case of H$_{2}$ in a circumstellar disk, we define a simple metric to characterize the average H$_{2}$ radius, $\langle$$R_{H2}$$\rangle$, 
\begin{equation}
\langle R_{H2} \rangle_{m} = GM_{*} \left (\frac{2 sin(i)}{FWHM_{m}} \right )^{2} 
\end{equation} 
where $M_{*}$ is the stellar mass, $i$ is the inclination angle, and FWHM$_{m}$ is the mean of the Gaussian FWHMs for a given progression $m$. This definition of the average molecular radius is analogous to the UV-CO radius used by~\citet{schindhelm11}. When possible, we use disk inclinations derived from the sub-mm dust continuum observations presented by~\citet{Andrews2007} and~\citet{Andrews2011}. In principle, the error on the average H$_{2}$ radius should include uncertainties on the stellar mass and disk inclinations, however these uncertainties are not available for all targets. Therefore, the quoted error on $\langle$$R_{H2}$$\rangle$ only includes measurement uncertainties from the H$_{2}$ line fitting. 

In Table 3 we present $\langle$$R_{H2}$$\rangle$ for the [$v^{'}$, $J^{'}$] = [1,7] progression ($\equiv$~$\langle$$R_{H2}$$\rangle$$_{[1,7]}$). We focus on the [1,7] progression due to its proximity to the Ly$\alpha$ line-center. This progression should be one of the most readily observable in weakly accreting systems because as the accretion-powered Ly$\alpha$ flux decreases, the flux in the wings of Ly$\alpha$ may be insufficient to excite a detectable level of H$_{2}$ emission.  However, the [1,7] progression should continue to be observable even in systems with narrow Ly$\alpha$ profiles (e.g., France et al. 2010).~\nocite{france10b} 
While we concentrate on the $\langle$$R_{H2}$$\rangle$$_{[1,7]}$, the $\langle$$R_{H2}$$\rangle$$_{[1,4]}$ distribution is qualitatively similar. We do not compute $\langle$$R_{H2}$$\rangle$ for targets with inclination angles $<$ 15\arcdeg; the small radial component of the H$_{2}$ velocity vector makes the derived radii very sensitive to uncertainties in the disk geometry. We also do not compute $\langle$$R_{H2}$$\rangle$ for the only unambiguously outflow-dominated source in the sample, RW Aur (\S4.2.1). The impact of a second, weaker outflow emission component will bias our results towards smaller H$_{2}$ radii in some targets, but a single component dominates the majority of our sources, therefore we adopt a single emission component in order to facilitate a uniform spectral analysis. 

A knowledge of the inner disk radius, $R_{in}$, is important for understanding the star-disk interaction. In order to avoid complications from outflows or blending from adjacent weaker H$_{2}$ transitions, we do not measure $R_{in}$(H$_{2}$) from half-width zero-intensities (HWZI, e.g., Brittain et al. 2009).\nocite{brittain09} Instead, we adopt the definition of $R_{in}$ suggested by~\citet{salyk11} for mid-IR CO emission from the inner disk, where $R_{in}$ is the Keplerian semi-major axis corresponding to 1.7~$\times$~HWHM of the Gaussian line fit. This choice leads to the relation between the average H$_{2}$ radius and the inner H$_{2}$ radius, $\langle$$R_{H2}$$\rangle$$_{[1,7]}$~=~2.89 $R_{in}$(H$_{2}$). 
$R_{in}$(H$_{2}$) is given with $\langle$$R_{H2}$$\rangle$$_{[1,7]}$ in Table 3.

\begin{figure}
\figurenum{8}
\begin{center}
\epsfig{figure=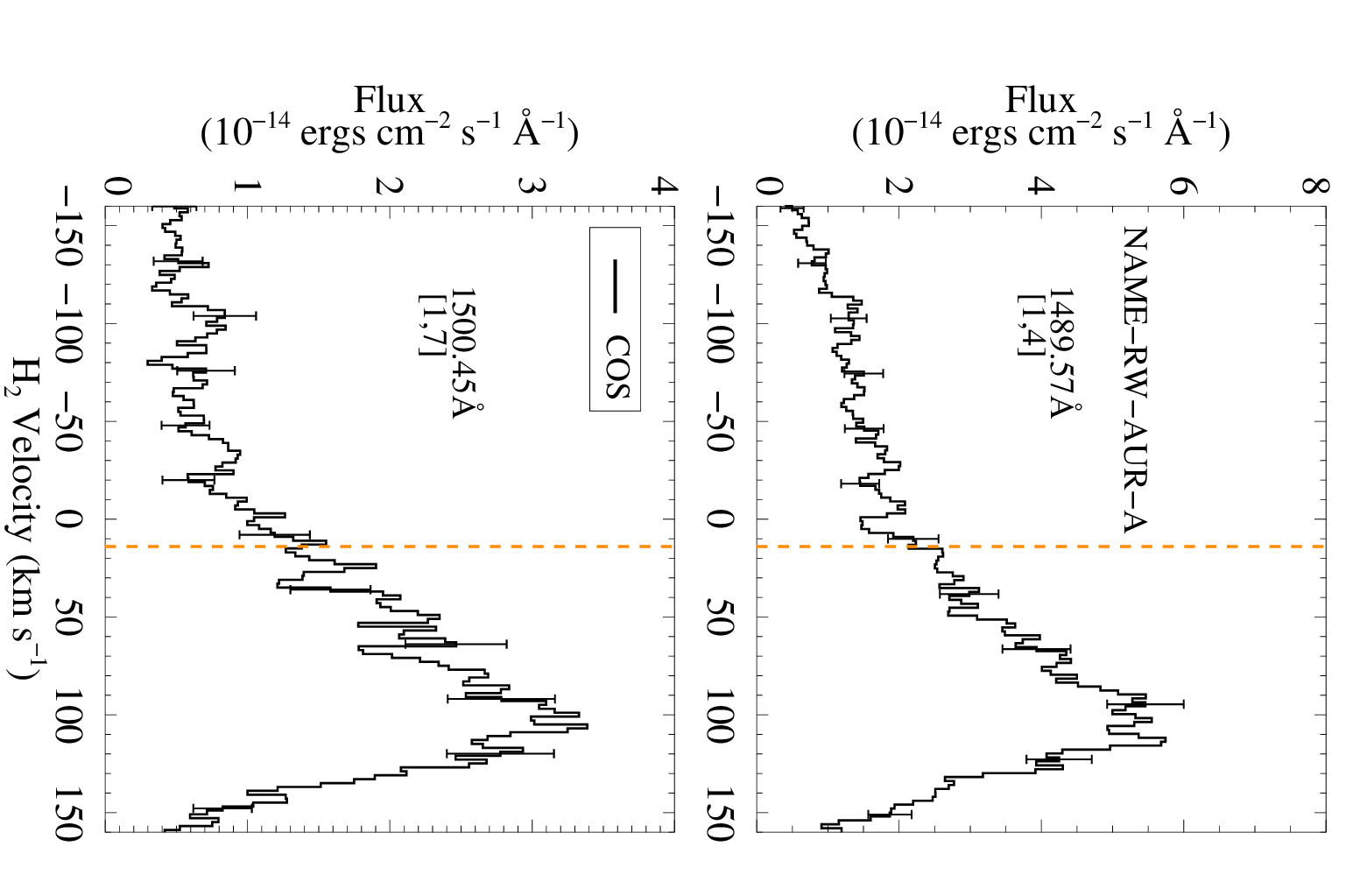,width=4.5in,angle=90}
\vspace{+0.15in}
\caption{
\label{cosovly} COS H$_{2}$ line profiles of RW Aur display a complicated kinematic behavior. 
Stellar variability compromises radial velocity measurements for RW Aur. We display +14 km s$^{-1}$, the canonical stellar radial velocity~\citep{hartmann86}, as the dashed orange line. 
The H$_{2}$ line profile is dominated by the redshifted outflow lobe~\citep{melnikov09} with an outflow velocity of $\approx$~100 km s$^{-1}$. 
}
\end{center}
\end{figure}

\begin{figure}
\figurenum{9}
\begin{center}
\epsfig{figure=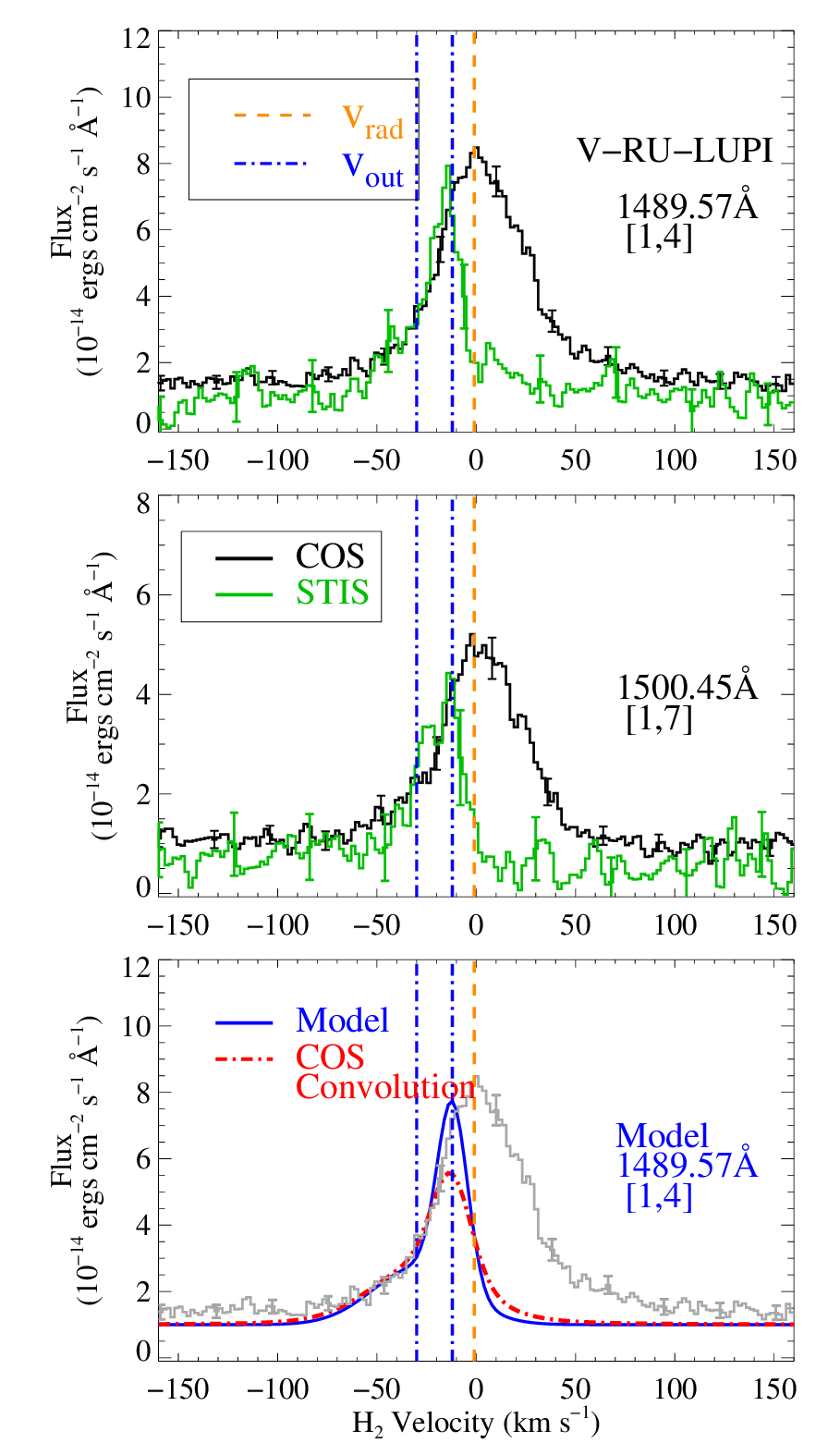,width=3.2in,angle=00}
\vspace{+0.15in}
\caption{
\label{cosovly} A comparison of archival STIS spectra of H$_{2}$ emission lines from RU Lupi~($green$) and new observations from COS ($black$). The stellar radial velocity is indicated with the orange dashed line and the outflow velocities from~\citet{herczeg06} are shown as the dash-dotted blue lines.
Strong lines from the [1,4] and [1,7] progressions are shown in the top two panels, and the bottom panel shows the two-component spectral fit for the STIS observations in blue~\citep{herczeg06}. This profile, convolved with the 1490~\AA\ COS LSF is shown as the red dash-dotted line and the observed COS data are shown in gray. The COS data are consistent with the stellar radial velocity and do not display asymmetric outflow profiles as strong as was observed by STIS. 
This difference may be due to a physical change in the inner regions of RU Lupi or could be explained by angular extension along the dispersion axis in the COS data (\S4.2.1). 
}
\end{center}
\end{figure}

\begin{deluxetable*}{lcccccccc}
\tabletypesize{\scriptsize}
\tablecaption{Molecular and Atomic Emission Line Parameters. \label{lya_lines}}
\tablewidth{0pt}
\tablehead{
\colhead{Target} & \colhead{FWHM$_{[1,7]}$\tablenotemark{a}} & \colhead{$\langle$$R_{H2}$$\rangle$$_{[1,7]}$\tablenotemark{a}} & 
\colhead{$R_{in}$\tablenotemark{b}} & \colhead{FWHM$_{CO}$\tablenotemark{c}} & \colhead{$n_{13-31}$} & \colhead{$L$(H$_{2}$)\tablenotemark{a}} & \colhead{$L$(Ly$\alpha$)\tablenotemark{d}} & \colhead{$L$(\ion{C}{4})\tablenotemark{e}} \\ 
& (km s$^{-1}$) & (AU) & (AU) & (km s$^{-1}$) & & (10$^{29}$ erg s$^{-1}$ ) & (10$^{29}$ erg s$^{-1}$ ) & (10$^{29}$ erg s$^{-1}$ )}
\startdata
AATAU & 62 $\pm$ 4 & 0.69 $\pm$ 0.08 & 0.24 & 92 & -0.51 & 46.7 $\pm$ 3.1 & 818.5 $\pm$ 146.6 & 6.1 \\ 
AKSCO & 57 $\pm$ 35 & 1.25 $\pm$ 0.77 & 0.43 & $\cdots$ & $\cdots $ & 8.1 $\pm$ 1.3 & 114.4 & 11.8 \\ 
BPTAU & 70 $\pm$ 6 & 0.13 $\pm$ 0.02 & 0.05 & 87 & -0.58 & 68.0 $\pm$ 7.4 & 731.7 $\pm$ 129.7 & 40.3 \\ 
CSCHA & 18 $\pm$ 7 & 9.00 $\pm$ 4.55 & 3.11 & $\cdots$ & 2.89 & 189.9 $\pm$ 12.9 & 2076.7 & 20.6 \\ 
CVCHA & 22 $\pm$ 30 & 4.75 $\pm$ 3.88 & 1.64 & $\cdots$ & -0.27 & 139.1 $\pm$ 35.5 & 1540.7 & 169.1 \\ 
DETAU & 55 $\pm$ 6 & 0.23 $\pm$ 0.04 & 0.08 & $\cdots$ & -0.12 & 20.1 $\pm$ 3.0 & 361.2 $\pm$ 106.0 & 7.8 \\ 
DFTAU\tablenotemark{f} & 64 $\pm$ 7 & 0.16 $\pm$ 0.03 & 0.06 & 79 & -1.09 & 95.7 $\pm$ 4.6 & 1064.7 & 23.2 \\ 
DKTAU \tablenotemark{f} & 55 $\pm$ 2 & 0.24 $\pm$ 0.02 & 0.08 & $\cdots$ & -0.81 & 21.3 $\pm$ 1.2 & 276.7 & 4.6 \\ 
DMTAU & 27 $\pm$ 5 & 0.80 $\pm$ 0.24 & 0.28 & $\cdots$ & 1.30 & 9.7 $\pm$ 0.7 & 106.5 $\pm$ 11.7 & 1.9 \\ 
DNTAU & 71 $\pm$ 19 & 0.09 $\pm$ 0.04 & 0.03 & $\cdots$ & -0.43 & 210.6 $\pm$ 67.8 & 2279.5 & 290.8 \\ 
DRTAU & 35 $\pm$ 7 & 2.09 $\pm$ 0.62 & 0.72 & 29 & -0.40 & 14117.7 $\pm$ 1590.0 & 149385.1 & 3986.7 \\ 
GMAUR & 41 $\pm$ 11 & 1.68 $\pm$ 0.65 & 0.58 & 47 & 1.76 & 18.5 $\pm$ 1.8 & 286.1 $\pm$ 70.4 & 7.6 \\ 
HD104237 & 94 $\pm$ 77 & 0.10 $\pm$ 0.07 & 0.03 & $\cdots$ & $\cdots $ & 964.4 $\pm$ 315.6 & 10239.6 & 91.6 \\ 
HD135344B & 26 $\pm$ 1 & $\cdots $ & $\cdots $ & 47 & $\cdots $ & 15.1 $\pm$ 1.3 & 212.0 & 27.4 \\ 
HNTAU\tablenotemark{f} & 61 $\pm$ 17 & 0.47 $\pm$ 0.18 & 0.16 & $\cdots$ & -0.44 & 19.1 $\pm$ 1.1 & 307.2 $\pm$ 60.3 & 2.6 \\ 
IPTAU\tablenotemark{f} & 102 $\pm$ 29 & 0.17 $\pm$ 0.07 & 0.06 & $\cdots$ & -0.11 & 4.0 $\pm$ 0.2 & 94.0 & 1.2 \\ 
LkCa15 & 53 $\pm$ 3 & 0.62 $\pm$ 0.06 & 0.21 & $\cdots$ & 0.62 & 24.4 $\pm$ 1.3 & 403.4 $\pm$ 66.4 & 9.2 \\ 
RECX11 & 54 $\pm$ 3 & 0.85 $\pm$ 0.08 & 0.29 & $\cdots$ & $\cdots $ & 1.9 $\pm$ 0.1 & 58.3 $\pm$ 3.9 & 1.7 \\ 
RECX15\tablenotemark{f} & 41 $\pm$ 4 & 0.62 $\pm$ 0.10 & 0.21 & $\cdots$ & $\cdots $ & 5.4 $\pm$ 0.3 & 130.6 $\pm$ 17.8 & 0.4 \\ 
RULUPI & 40 $\pm$ 2 & 0.30 $\pm$ 0.03 & 0.10 & 24 & $\cdots $ & 27.2 $\pm$ 15.2 & 635.3 $\pm$ 148.3 & 12.0 \\ 
RWAUR\tablenotemark{g} & $\cdots $ & $\cdots $ & $\cdots $ & $\cdots$ & -0.54 & 860.7 $\pm$ 228.9 & 9169.5 & 58.9 \\ 
SUAUR & 49 $\pm$ 6 & 2.67 $\pm$ 0.58 & 0.92 & 121 & 0.74 & 36.3 $\pm$ 4.0 & 811.4 $\pm$ 191.4 & 17.3 \\ 
SZ102 & 47 $\pm$ 7 & $\cdots $ & $\cdots $ & $\cdots$ & $\cdots $ & 196.2 $\pm$ 13.8 & 2182.4 & 60.8 \\ 
TWHYA & 18 $\pm$ 2 & $\cdots $ & $\cdots $ & 17 & $\cdots $ & 16.8 $\pm$ 2.0 & 199.6 $\pm$ 34.3 & 17.0 \\ 
UXTAU & 29 $\pm$ 3 & 1.76 $\pm$ 0.33 & 0.61 & 21 & 1.83 & 12.6 $\pm$ 0.3 & 146.3 $\pm$ 12.3 & 2.4 \\ 
V4046SGR & 45 $\pm$ 1 & 0.95 $\pm$ 0.06 & 0.33 & $\cdots$ & $\cdots $ & 19.8 $\pm$ 0.9 & 383.3 $\pm$ 42.1 & 4.5 \\ 
V836TAU & 47 $\pm$ 20 & 0.99 $\pm$ 0.50 & 0.34 & $\cdots$ & -0.45 & 80.2 $\pm$ 7.3 & 900.5 & 17.7 \\ 
\tableline
\tableline
HBC427\tablenotemark{h} & $\cdots $ & $\cdots $ & $\cdots $ & $\cdots$ & $\cdots $ & $\leq$ 0.5 & $\leq$ 57.7 & 0.2 \\ 
LKCA19 & $\cdots $ & $\cdots $ & $\cdots $ & $\cdots$ & $\cdots $ & $\leq$ 1.1 & $\leq$ 63.6 & 0.5 \\ 
LKCA4 & $\cdots $ & $\cdots $ & $\cdots $ & $\cdots$ & $\cdots $ & $\leq$ 1.0 & $\leq$ 62.6 & 1.6 \\ 
RECX1 & $\cdots $ & $\cdots $ & $\cdots $ & $\cdots$ & $\cdots $ & $\leq$ 0.3 & $\leq$ 27.7 & 0.5 \\ 
TWA13A & $\cdots $ & $\cdots $ & $\cdots $ & $\cdots$ & $\cdots $ & $\leq$ 0.1 & $\leq$ 9.2 & 0.1 \\ 
TWA13B & $\cdots $ & $\cdots $ & $\cdots $ & $\cdots$ & $\cdots $ & $\leq$ 0.2 & $\leq$ 10.2 & 0.1 \\ 
TWA7 & $\cdots $ & $\cdots $ & $\cdots $ & $\cdots$ & $\cdots $ & $\leq$ 0.0 & $\leq$ 8.6 & 0.1  
\enddata
\tablenotetext{a}{Average FWHM and average H$_{2}$ radius (see \S3.2) were calculated from four lines of the H$_{2}$ $B$$^{1}\Sigma^{+}_{u}$~--~$X$$^{1}\Sigma^{+}_{g}$ (1~--~$v^{''}$) R(6)+P(8) progression. The H$_{2}$ luminosity is the sum of the 12 progressions measured in this work (Table 2).} 
\tablenotetext{b}{Inner H$_{2}$ radius, defined as $R_{in}$ = $GM_{*}$ (sin($i$)/(1.7~$\times$~HWHM$_{[1,7]}$))$^{2}$. } 
\tablenotetext{c}{CO line-widths taken from~\citet{salyk11} and \citet{bast11}. } 
\tablenotetext{d}{\ion{H}{1} Ly$\alpha$ luminosities with error bars were calculated from the Ly$\alpha$ fluxes presented by~\citet{schindhelm12a}. Values without error bars were extrapolated from the relationship between $F$(H$_{2}$) and $F$(Ly$\alpha$). } 
\tablenotetext{e}{Measurement error on the \ion{C}{4} flux is taken as 5\%.} 
\tablenotetext{f}{Targets where molecular outflows may contribute to the observed H$_{2}$ line-width. } 
\tablenotetext{g}{Strong molecular outflows in RW Aur contaminate the Gaussian fitting (see Figures 3$b$ and 8). } 
\tablenotetext{h}{Targets below the double line do not show measurable H$_{2}$ emission in their far-UV spectra. } 
\end{deluxetable*}


\section{Discussion}

There are numerous indicators of the gas and dust content of a young protoplanetary system. Three important observables are the warm dust content of the inner disk, the presence of circumstellar gas, and signs of active accretion. Our $HST$ observations provide measurements of the last two, while the first has been extensively studied in the IR. The combination of low spectral resolution and large instrumental backgrounds that have complicated the detection and analysis of H$_{2}$ in previous UV surveys has largely been remedied with the installation of COS. The large transition probabilities of the H$_{2}$ electronic band systems and the lack of photospheric emission at $\lambda$~$<$~1700~\AA\ in low-mass stars make fluorescent H$_{2}$ one of the most sensitive indicators for the presence of molecular gas in the inner ~$\sim$~10~AU of young circumstellar disks. 

Accretion shocks are a significant source of hot gas in accreting systems, observed as UV and X-ray emission lines in excess of what can be attributed to magnetospheric activity alone~\citep{calvet98,krull00,gunther08}. Specifically, excess emission from neutral hydrogen (line formation temperature $T_{form}$~$\sim$~10$^{4}$ K; observed as Ly$\alpha$ and H$\alpha$ emission) and the C$^{3+}$ ion ($T_{form}$~$\sim$~10$^{5}$ K; observed through the $\lambda\lambda$~1548, 1550~\AA\ \ion{C}{4} resonance doublet) correlate well with both the mass accretion rate and the H$_{2}$ emission from the system~\citep{krull00}. This supports a symbiotic picture where gas-rich disks provide fuel for active accretion, and that accretion dominates the production of the Ly$\alpha$ photons that make the H$_{2}$ disk detectable. 

The total H$_{2}$ luminosity is compared with the Ly$\alpha$ and \ion{C}{4} luminosities in Figure 6. The general trend follows the expected relation that systems with larger $L$(Ly$\alpha$) and $L$(\ion{C}{4}) are actively accreting gas-rich disks. 
Ly$\alpha$ fluxes are only available for about half of our sample (see Schindhelm et al. 2012b), and we interpolate (or extrapolate) the strong correlation between $F$(Ly$\alpha$) and $F$(H$_{2}$) to estimate $L$(Ly$\alpha$) for the remaining objects. 
The direct measurements are shown as black squares in Figure 6, the interpolated (extrapolated) values are shown in green, and upper limits are indicated in red. 
The correlation between $L$(H$_{2}$) and $L$(\ion{C}{4}) has a spread of $\sim$~1~--~1.5 orders of magnitude in $L$(H$_{2}$) at a given \ion{C}{4} luminosity. This spread may be partially due to uncertainties in the distance and reddening correction used to derive the luminosities, disks/outflows with differing amounts of molecular gas, and intrinsic variations in the \ion{C}{4} flux.
In the following subsections, we combine our molecular and atomic tracers to constrain the properties of H$_{2}$ in the circumstellar environments of these systems. 

\begin{figure}
\figurenum{10}
\begin{center}
\epsfig{figure=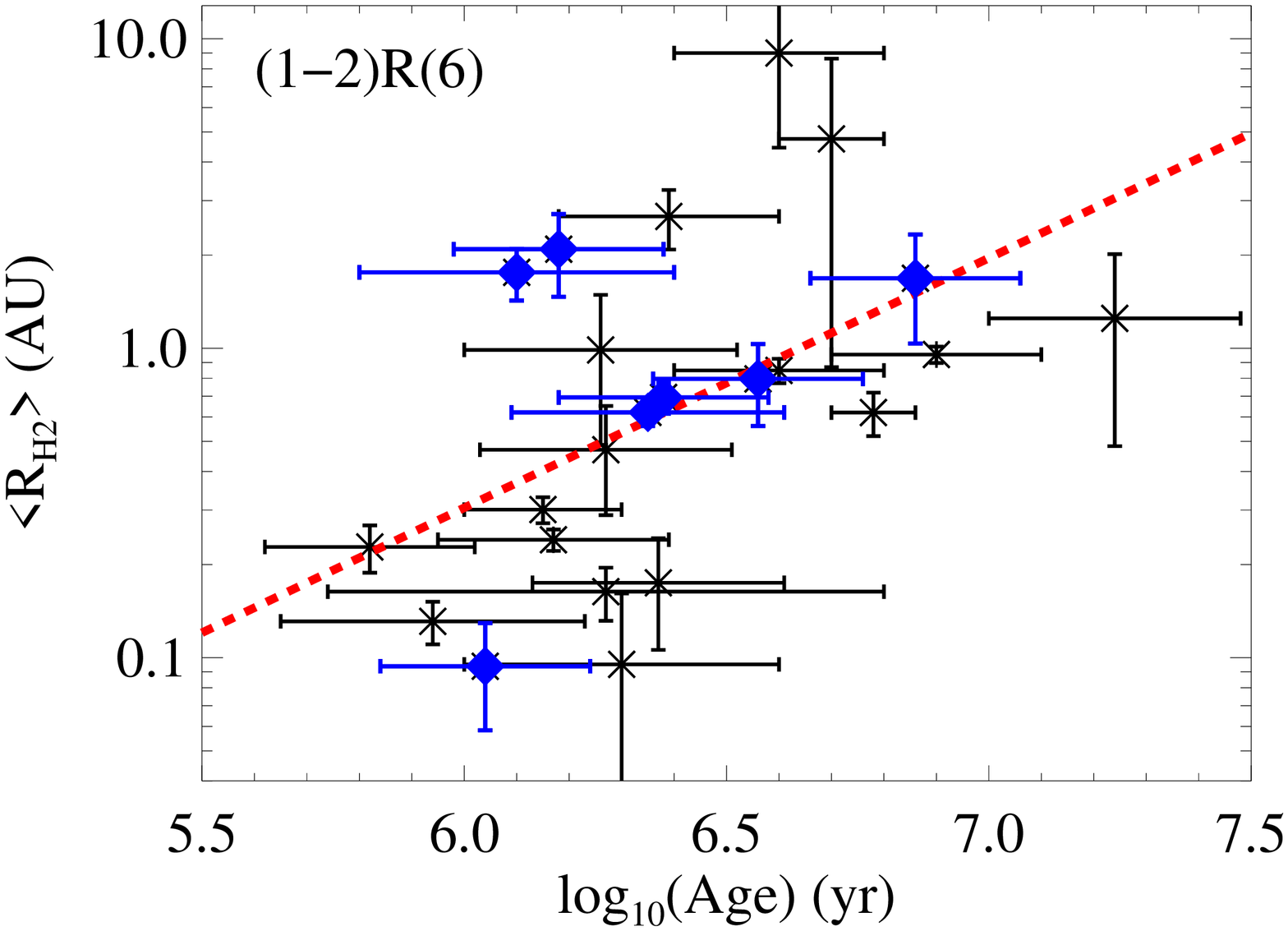,width=2.5in,angle=90}
\epsfig{figure=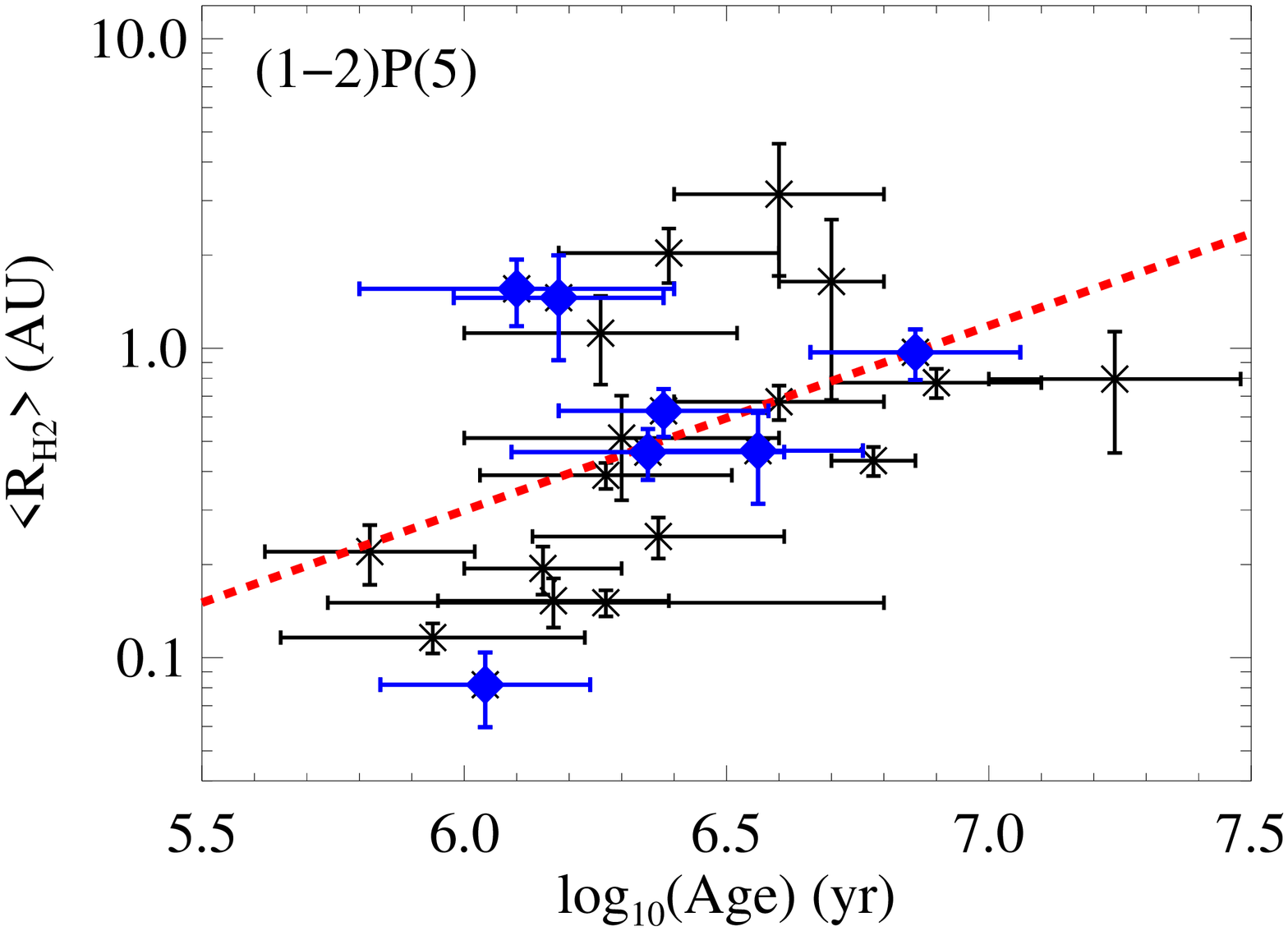,width=2.5in,angle=90}
\vspace{-0.1in}
\caption{
\label{cosovly} The time evolution of the average radial position of the H$_{2}$ emission in our sample, as defined by the Gaussian line-widths of the [1,7] (pumped through the (1~--~2) R(6) absorption line) and [1,4] (pumped through the (1~--~2) P(5) absorption line) progressions (see \S3.2 for a discussion). Only targets with disk inclinations $i$~$>$~15\arcdeg\ were considered. The red dashed line represents the empirical relation between the average radius of the H$_{2}$ emitting disk and the system age (Equation 3) for the [1,7] and [1,4] progressions. The two H$_{2}$ progressions show a similar behavior with system age. The Spearman rank correlation coefficient for increasing molecular radius with system age for both H$_{2}$ progressions is~$\approx$~0.52. The blue diamonds indicate targets with inclinations derived from sub-mm dust continua~\citep{Andrews2007,Andrews2011}. 
}
\end{center}
\end{figure}

\subsection{Evolution of the H$_{2}$ Luminosity and Reprocessing of the Ly$\alpha$ Radiation Field}

The dissipation of inner dust disks (at $a$~$<$~1AU) is thought to be mostly complete by $\approx$~6 Myr (e.g., Haisch et al. 2001; Wyatt 2008 and references therein). If the molecular gas and dust are coupled, we would expect to observe a decrease in the H$_{2}$ content as a function of system age. The evolution of the H$_{2}$ luminosity is plotted in the top panel of Figure 7. The H$_{2}$ detections are shown in black and the non-detections in red. The total H$_{2}$ luminosity decreases with time, in agreement with the findings of~\citet{ingleby09,ingleby11b}. 
However, the large scatter in the relation prevents one from inferring a characteristic timescale for this decrease. The dominant sources of uncertainty in this relation are the correlated errors on the age and extinction. 
The observed decrease in $L$(H$_{2}$) as a function of time does not necessarily correspond to a monotonic decrease in the H$_{2}$ content. We expect that the Ly$\alpha$ luminosity should track the accretion rate (e.g. Fang et al. 2009 demonstrate the correlation between accretion and H Balmer emission), therefore as the accretion rates decline over time~\citep{armitage03,aguilar10}, the associated reduction in Ly$\alpha$ emission could mimic the effect of H$_{2}$ dissipation with age. The true situation is likely a combination of these effects: gas disk dissipation leads to lower accretion rates that in turn produce fewer Ly$\alpha$ photons to pump the observed fluorescence. \nocite{fang09}

The lower panel in Figure 7 shows the ratio of the H$_{2}$ flux and the Ly$\alpha$ flux as a function of system age for the 14 objects 
with computed Ly$\alpha$ fluxes~\citep{schindhelm12a}. 
This plot shows the fraction of the stellar + shock Ly$\alpha$ that is reprocessed by H$_{2}$. The actual reprocessing factor is $\eta$$F$(H$_{2}$)/$F$(Ly$\alpha$), where $\eta$ is a correction factor to account for anisotropies in the system geometry and fluorescent radiative transfer~\citep{wood02,wood04}. In the case of isotropic emission, $\eta$ is simply the geometric filling fraction of the H$_{2}$, as seen by the Ly$\alpha$ photons. For this comparison (and for the computation of $L$(Ly$\alpha$) in Figure 6), we have assumed $\eta$~=~1. For sources where the majority of the H$_{2}$ resides in a flattened disk, $\eta$ is most likely less than one (Herczeg et al. 2004 found $\eta$~$\approx$~0.25 in model fits to the spectrum of TW Hya), while for sources with a significant outflow component $\eta$ may be $\sim$~1. 
\citet{yang11} and \citet{france12a} present analyses of H$_{2}$ absorption lines imposed on the Ly$\alpha$ profiles of the CTTSs V4046 Sgr, DF Tau, and AA Tau, systems with both high and low inclinations.  This implies that at least some portion of the H$_{2}$ in these systems has a Ly$\alpha$ covering fraction of near unity, suggesting that $\eta$~$\sim$~1 even in some disk-dominated systems. 
Furthermore, in cases where Ly$\alpha$ has been scattered out of our line of sight or self-absorption redistributes the fluorescence to the higher $v^{''}$ levels we use to determine $F$(H$_{2}$), $\eta$ can be $>$~1~\citep{wood02}. In the absence of a more sophisticated radiative transfer treatment of each system individually, we assume $\eta$~=~1 for the present analysis. 

The total Ly$\alpha$ flux is the full, unabsorbed Ly$\alpha$ profile as it is emitted from the immediate stellar environment. 
Figure 7 shows that the ratio of the total emitted Ly$\alpha$ flux that is reprocessed by H$_{2}$ is 6.2~$\pm$~2.1\%. 
A more meaningful measure of the degree of H$_{2}$ reprocessing is the ratio of incident Ly$\alpha$ that arrives at the molecular material. The Ly$\alpha$ profile will experience some degree of absorption in the circumstellar environment prior to reaching the molecules~\citep{wood04,herczeg04,schindhelm11}. 
Using the incident Ly$\alpha$ profile observed by the H$_{2}$, we find that the reprocessing fraction ($\eta$$F$(H$_{2}$)/$F$(Ly$\alpha$)) is 11.5~$\pm$~1.8\%.
Therefore, modulo the factor of $\eta$, we infer that H$_{2}$ is capable of reprocessing~$>$~10\% of the incident Ly$\alpha$ flux. This is interesting because the H$_{2}$ $a)$ will isotropically redistribute the Ly$\alpha$ photons and $b)$ represents a means for transferring Ly$\alpha$ photons out of the Ly$\alpha$ line-core and redistributing them across the 1000~$\lesssim$~$\lambda$~$\lesssim$~1650~\AA\ bandpass.  Additionally, H$_{2}$ scattering is a means for redirecting Ly$\alpha$ photons initially on a grazing incidence trajectory, increasing the far-UV radiation penetration depth.  This will significantly alter the radiative transfer of this fraction of the Ly$\alpha$ energy as it diffuses outward and downward through the disk~\citep{fogel11,bethell11}. The transfer of these H$_{2}$-redistributed Ly$\alpha$ photons will be regulated by circumstellar grains, and this process will add power to discrete wavelengths in the far-UV spectrum that propagates towards the disk midplane, possibly perturbing disk chemistry in regions of active planet formation.


\subsection{Spatial Distribution of H$_{2}$}

\subsubsection{H$_{2}$ Outflows} 

A single Gaussian emission line describes many of the observed velocity profiles, however several targets show evidence for molecular outflows in the form of additional red/blue-shifted H$_{2}$ emission. \citet{beck08} presented an outflow-selected sample of CTTSs that display spatially extended near-IR rovibrational H$_{2}$ spectra.
Furthermore, \citet{pontoppidan11} have found slow (5~--~10 km s$^{-1}$), weakly collimated molecular winds to be common in CO spectra of CTTSs, and varying contributions from these winds/outflows could be responsible for the blue-shifted H$_{2}$ that is observed towards some systems. 
The COS observations of DF Tau, DK Tau, HN Tau, LkCa15, IP Tau, RECX-15, RU Lupi, and RW Aur all display H$_{2}$ line wings extending to the blue of the stellar radial velocity. Interestingly, half of these objects are known binaries (DF Tau, DK Tau, HN tau, and RW Aur), and this may indicate that interactions with a companion star contribute to the generation of molecular outflows in CTTSs.  Clearly, more data is required to test this connection.  

Of the targets displaying H$_{2}$ emission with extended blue wings, DF Tau, DK Tau, HN Tau, IP Tau, RECX-15, and RW Aur also have reasonably high S/N [0,1] progression emission lines in their spectra. The [0,1] line-widths are statistically narrower than those from [1,7] and [1,4], also suggesting that outflow contributions to the [1,4] and [1,7] progressions are present in these targets~\citep{walter03}. 
RW Aur shows the strongest outflow emission in the survey, with both red- and blue-shifted emission observed. It is not clear that there is a narrow disk H$_{2}$ component present in this source. We show a blow-up of bright lines from the [1,7] and [1,4] progressions in RW Aur in Figure 8. RW Aur is known to have a bipolar outflow, with the red component ($v_{out}$~$\sim$~+100 km s$^{-1}$) being brighter and higher density than the blue~\citep{hirth94,melnikov09}. Due to the COS aperture vignetting function, the observed spectra will be dominated by the inner $\approx$~1\arcsec\ of the RW Aur jet. Figure 8 shows that the H$_{2}$ emission peaks at an observed velocity 80~--~110 km s$^{-1}$ to the red of the stellar velocity (+14 km s$^{-1}$; Hartmann et al. 1986), suggesting that the molecular emission arises in material that is approximately cospatial with the forbidden atomic line (e.g., [\ion{S}{2}] $\lambda$6731 \AA) emission~\citep{woitas02,melnikov09,hartigan09}. The near-IR H$_{2}$ outflow from RW Aur is centered near $\sim$~+44 km s$^{-1}$~\citep{beck08}, significantly bluer than the peak of the far-UV H$_{2}$ velocity profile. It is not clear if this indicates a difference in the physical structure of the UV and IR-emitting H$_{2}$, or can be attributed to blending of low- and high-velocity gas in the lower spectral resolution near-IR data. 
For the remaining targets, the blue-wings are relatively weak, typically only perturbations from the narrower, presumably disk-dominated H$_{2}$ velocity profiles. Further study using coadded spectra from several progressions would be useful for clarifying the outflow contribution and structure in these targets. 

The observed H$_{2}$ velocity profiles of RU Lupi are puzzling. Using higher spectral resolution observations from STIS, \citet{herczeg05,herczeg06} found that essentially all of the fluorescent H$_{2}$ was contained in two blue-shifted components, ~--~12 and ~--~30 km s$^{-1}$ relative to the radial velocity of the star.\nocite{herczeg05} While a blue-wing is apparent in the COS observations, the emission is well-fit by a single component at the radial velocity of the star. In Figure 9, we compare the COS observations (obtained in 2011 using the 2.5\arcsec\ diameter Primary Science Aperture) with the STIS observations (obtained in 2000 using the 0.2\arcsec~$\times$~0.06\arcsec\ slit). The apparent line-center velocity shift is $\approx$~15~--~20 km s$^{-1}$, larger than the zero-point calibration of the COS wavelength solution\footnote{Calibration of the COS wavelength solution is limited by systematic uncertainty in the far-UV detector geometric correction.}. Furthermore, the line-shape is fundamentally different in a way that cannot be explained by resolution differences between the two instruments. In the bottom panel of Figure 9, we reconstruct a model two-Gaussian profile (solid blue line) using the fit parameters from~\citet{herczeg06}. That profile, convolved with the COS LSF at 1490~\AA~\citep{kriss11}, is displayed as the red dash-dotted curve. CTTSs (and in particular RU Lupi; Gahm et al. 2008) are known to be time-variable, therefore it is not surprising that the peak flux has changed, but the line center and emission line shapes are not consistent.\nocite{gahm08} 

\citet{herczeg06} found that the H$_{2}$ fluorescence in RU Lupi was spatially extended,  although the small size 
of the STIS aperture makes it difficult to predict the effects of spatial extension in the larger COS aperture.  
An aperture offset of~$\gtrsim$~0.4\arcsec\ would be required for significant reduction in the instrumental resolving power, but angular extension of the H$_{2}$ emitting region may be able to alter both the velocity centroid and the line-width.   In the dispersion direction, if we attribute the entire 20 km s$^{-1}$ offset to extended emission, this leads to a 0.2\arcsec\ displacement (at the 24.3 milliarcseconds pixel$^{-1}$ dispersion-direction plate-scale of the G160M mode). 
We cannot constrain the angular extent of the RU Lupi H$_{2}$ lines in the dispersion direction, but we note that the optical forbidden line emission ([\ion{O}{1}] and [\ion{S}{2}]) are extended to~$\sim$~0.2\arcsec\ at a position angle of~$\sim$~225\arcdeg~\citep{takami01}.  If the H$_{2}$ emission lines observed in the COS spectra are cospatial with the forbidden line emission, then this could produce the $\sim$ + 20 km s$^{-1}$ velocity offset.  Therefore, it is possible that spatial extension of the H$_{2}$ emitting gas along this axis could explain the differences in line shape between the STIS and COS observations.  We compared the cross-dispersion profile of the two-dimensional spectrogram of the RU Lupi observations (over the range $\Delta$$\lambda$~$\sim$~1420~--~1450~\AA) with that of the DA white dwarf WD0320-539. The two spectra were centered to within 0.5 pixels ($\approx$~0.05\arcsec) in the cross-dispersion direction and had nearly identical profile FWHMs (4.5 pixels for WD0320, 4.8 pixels for RU Lupi). 

We conclude therefore that either RU Lupi has spatial extent along the dispersion axis, or the spectral profile differences between the COS and STIS epochs are caused by a physical change in the system. RU Lupi may continue to be outflow dominated, however the COS observations raise the possibility that that RU Lupi was observed during an episode of strong outflow in 2000 or that the geometry has evolved such that disk-illumination by Ly$\alpha$ contributed more strongly during the 2011 observations. Continued spectral monitoring of these loopy line profiles would be interesting. 

\begin{figure}
\figurenum{11}
\begin{center}
\epsfig{figure=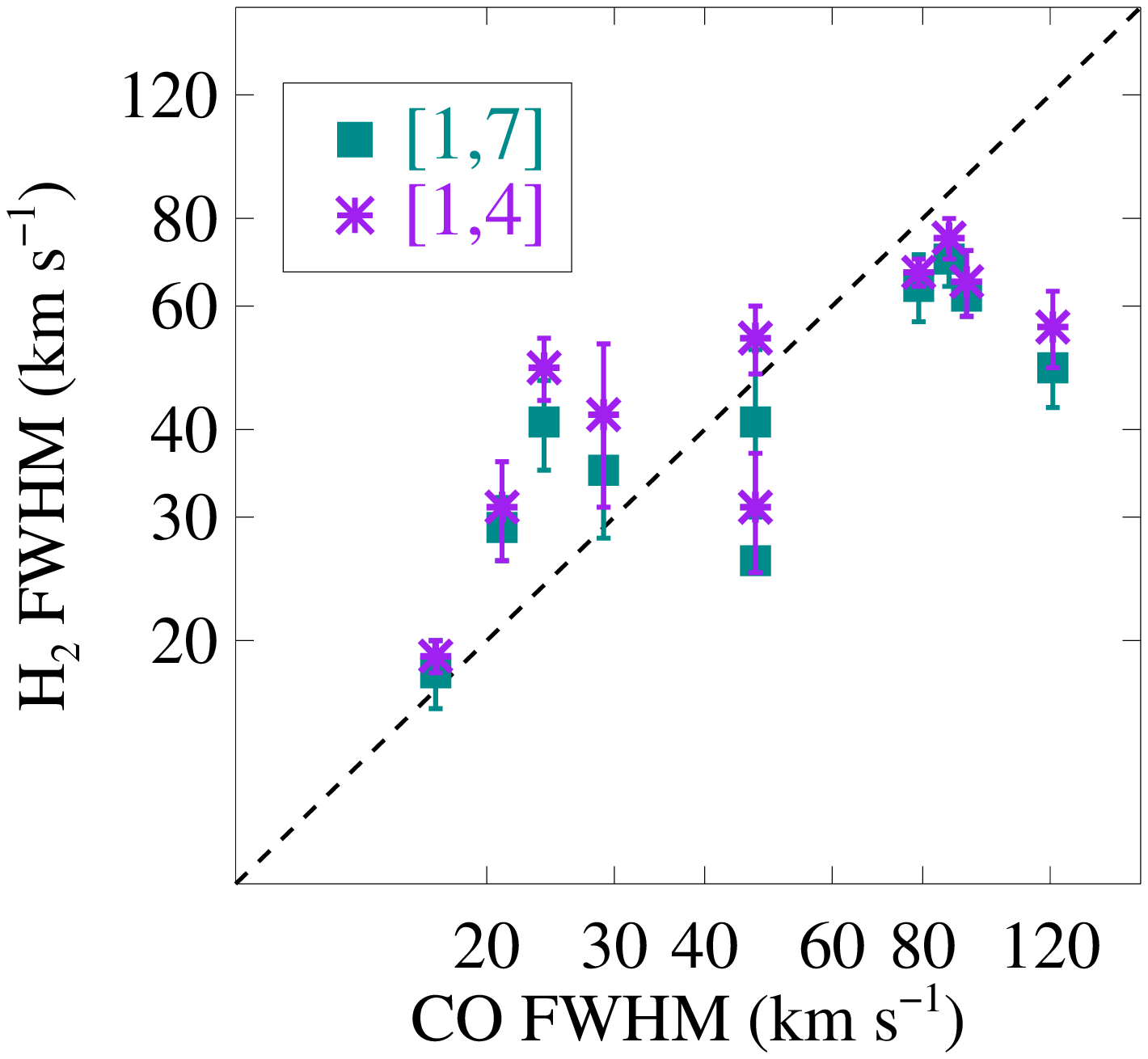,width=3.0in,angle=90}
\vspace{-0.2in}
\caption{
\label{cosovly} A comparison of the emission line FWHMs of H$_{2}$ (this work) and CO~\citep{salyk11,bast11}. 
The squares are the Gaussian line-widths of the [1,7] progression and the stars are the Gaussian line-widths of the [1,4] progression. The dashed line represents the expected correlation if the H$_{2}$ and CO line-widths are the same. 
}
\end{center}
\end{figure}

\begin{figure}
\figurenum{12}
\begin{center}
\epsfig{figure=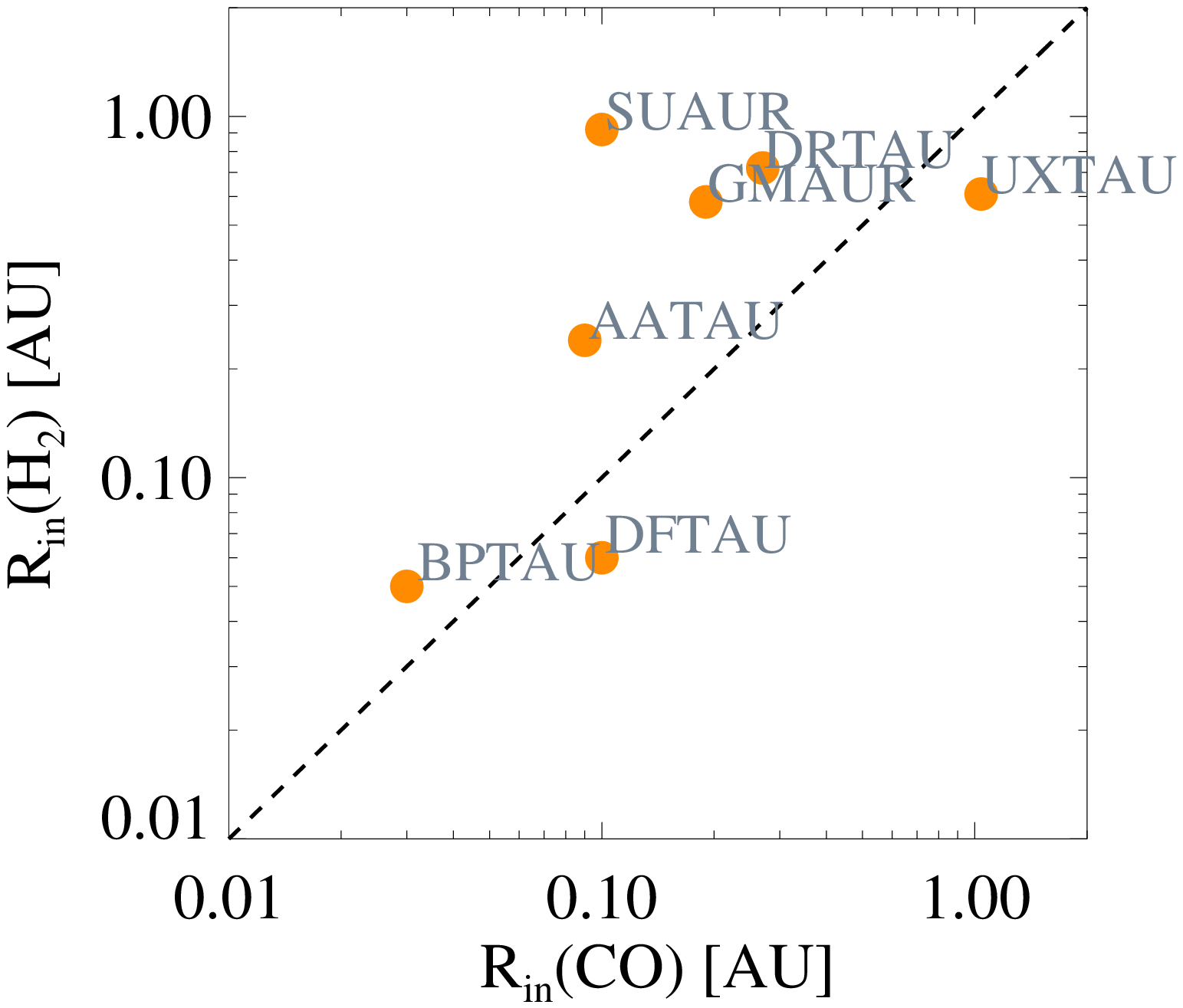,width=3.0in,angle=90}
\vspace{-0.2in}
\caption{
\label{cosovly} A comparison of inner radii ($R_{in}$) of the molecular disk traced by H$_{2}$ (this work) and CO~\citep{salyk11}. The dashed line represents the expected correlation if the H$_{2}$ and CO emission are co-spatial. 
$R_{in}$ is defined as the Keplerian radius corresponding to an orbital velocity of 1.7~$\times$~HWHM of the Gaussian line fits (\S3.2). $R_{in}$(H$_{2}$) is computed from the [1,7] progression for this comparison. 
Except in the case of SU Aur, $R_{in}$(H$_{2}$) and $R_{in}$(CO) are the same to within a factor of three, suggesting that the observed CO and H$_{2}$ orbit at similar radii.
}
\end{center}
\end{figure}

\subsubsection{H$_{2}$ Disks} 

The high S/N H$_{2}$ lines in the majority of our sample targets can be adequately described by a single Gaussian emission component at the stellar radial velocity. Bearing in mind the added uncertainty introduced by blue-shifted H$_{2}$ in some targets, we conclude that most of the observed fluorescent emission originates in a disk. This is in contrast with the near-IR H$_{2}$ study of~\citet{beck08}; however, that study targeted stars known to have strong outflows, which is not the case with the sample presented here.   Using the velocity-resolved H$_{2}$ line profiles, $\langle$$R_{H2}$$\rangle$ was calculated for the [1,7] and [1,4] progressions. The radial distributions as a function of system age are displayed in Figure 10. The general result is that the average location of the emitting H$_{2}$ gas in the inner disk moves outward from $\sim$~0.25 AU to $\sim$~2 AU as the system evolves from 1~--~10 Myr. 

The relationship between the system age and the average H$_{2}$ radius can be characterized by the simple empirical formula 
\begin{equation}
log_{10}\langle R(H_{2}) \rangle_{[1,7]}~=~0.81~log_{10}(Age) - 5.36
\end{equation}
where $\langle$$R_{H2}$$\rangle$$_{[1,7]}$ is in AU and the age is in years. The coefficients in Equation 3 [$-$5.36~$\pm$~1.91, 0.81~$\pm$~0.30] are computed from a $\chi^{2}$ minimization of a linear function of log$_{10}$(Age) and log$_{10}$$\langle$$R_{H2}$$\rangle$$_{[1,7]}$. The Spearman rank correlation coefficient for increasing molecular radius with system age is 0.52 with a deviation from zero of 1.1~$\times$~10$^{-2}$, 
meaning that there is a relatively low probability that log$_{10}$(Age) and log$_{10}$$\langle$$R_{H2}$$\rangle$$_{[1,7]}$ are uncorrelated. Targets with disk inclinations derived from sub-mm dust continuum observations~\citep{Andrews2007,Andrews2011} are plotted as the blue diamonds in Figure 10. There is a large spread in the distribution of $\langle$$R_{H2}$$\rangle$ with age. We suggest that in addition to contamination by outflows, uncertainties in the stellar mass, inclination angle, and stellar age all contribute to the dispersion in this correlation. With the exception of CS Cha (\S4.3), the H$_{2}$ in all disks is concentrated at $a$~$\lesssim$~3 AU. 

The weak trend towards larger radii as a function of age suggests a scenario where the average molecular emission radius moves to beyond $\approx$~1 AU in about~10~Myr. It is not immediately clear how to interpret this result, and we remind the reader of the ingredients necessary to produce H$_{2}$ fluorescence in a disk. The H$_{2}$ 
opacity must be high enough to absorb a significant number of Ly$\alpha$ photons, requiring both appreciable column densities and a sufficient population of H$_{2}$ in excited ro-vibrational states. The latter requires a hot ($T$(H$_{2}$)~$\gtrsim$~2000 K) molecular layer, possibly with a contribution by intense illumination from the $\lambda$~$\leq$~1120~\AA\ continuum from the central star + accretion shocks~\citep{nomura07,france12a}. 
The second major requirement for observable fluorescence is a geometry where the disk subtends a substantial angular cross-section 
of the Ly$\alpha$ emitting area, that is, the disk must be sufficiently flared in order to intercept enough Ly$\alpha$ photons to excite the observed emission. The observation of larger H$_{2}$ radii as a function of time suggests that one or more of these criteria are not being met in the inner $\approx$~1 AU in the more evolved systems. It may be that the hot H$_{2}$ is in the process of dissipating from this region, possibly due to dynamical clearing by a protoplanet, enhanced H$_{2}$ dissociation as this region is less shielded by grains, or photoevaporation by energetic radiation from the central star. Alternatively, this result may indicate that the flaring angle in the inner disk is decreasing across our sample, suggesting an evolution of the vertical structure of the disk on timescales of a few Myr. Improved stellar masses and ages, and larger samples of well-determined disk inclinations would allow better characterization of the relation (or lack thereof) between molecular radius and age, enabling a better understanding of the evolution of inner gas disks. 

\subsubsection{Comparison with near-IR H$_{2}$ Emission} 

Quadrupole rovibrational line emission from H$_{2}$ (most notably the (1~--~0) S(1) $\lambda$2.1~$\mu$m line) has been detected around several CTTSs~\citep{bary03,itoh03,carmona07,bary08}.  While the number of objects available for direct comparison is small, it seems that the Ly$\alpha$-pumped H$_{2}$ emission arises interior to the near-IR emission lines (although see \S4.3 for the case of the transitional disk CS Cha). 
Typical emission line widths for the near-IR H$_{2}$ sample presented in~\citet{bary08} are FWHM~$\leq$~20 km s$^{-1}$, and are thought to originate at radial distances of a few to a few tens of AU.  Similarly, \citet{bary03} observed (1~--~0) S(1) emission in LkCa15 with FWHM~$\leq$~14 km s$^{-1}$ and suggested that the emitting gas was located between 10~--~30~AU from the star.  This is a factor of $\approx$~4 narrower than the FWHM$_{[1,7]}$~=~53~$\pm$~3 km s$^{-1}$ that we observe in the LkCa15 UV spectrum. 

RECX-15 is the only disk in the $\eta$ Cha region to emit a measurable flux of H$_{2}$ in the (1~--~0) S(1) line~\citep{ramsay07}. The near-IR H$_{2}$ line widths (18~$\pm$~1.2 km s$^{-1}$) are a factor of $\approx$~2.3 smaller than measured in the COS spectra of RECX-15 (41~$\pm$~4 km s$^{-1}$), although we note that outflows do contribute to the RECX-15 ultraviolet H$_{2}$ spectra. The near-IR H$_{2}$ emission from disks is typically interpreted as disk surface gas excited by energetic radiation. Including the effects of grain grown, UV, and X-ray illumination, \citet{nomura07} have demonstrated that gas temperatures in the range 1500~--~3000 K can be maintained in the disk surface to radial distances of $\approx$~10 AU from the central star, therefore the excitation conditions necessary to both produce near-IR emission and to enable Ly$\alpha$-pumping appear to exist to at least this radius. However, the fluorescent ultraviolet emission is dominated by H$_{2}$ nearest to the source of Ly$\alpha$ photons. 

\subsubsection{Comparison with CO Emission}

Comparing the line-widths of different molecular species can provide an observational constraint on the composition and physical structure of inner gas disks. Spectral observations of the 4.7~$\mu$m fundamental band CO emission are a widely used tracer for this material~\citep{salyk08,salyk09}, and understanding the molecular structure and the degree to which various spectral diagnostics trace the same gas are useful towards a more complete picture of the planet-forming regions around CTTSs. Figure 11 shows a comparison of the Gaussian FWHMs of CO and H$_{2}$ for the subsample of our targets that have been observed by high-resolution mid-IR spectrographs~\citep{salyk11, bast11}. The emission line-widths from the [1,7] and [1,4] H$_{2}$ progressions are self-consistent in all cases and approximately equal the CO line-widths up to FWHM~$\approx$~60 km s$^{-1}$. At larger CO line-widths, the H$_{2}$ FWHMs do not exceed 70~--~80 km s$^{-1}$, which may indicate a physical boundary condition inside of which H$_{2}$ is subject to collisional and/or photodissociation. 

The inner radii of the H$_{2}$ and CO disks can also be directly compared. The calculated H$_{2}$ inner radii are presented in Table 3, and Figure 12 compares the ultraviolet H$_{2}$ (UV-H$_{2}$) and infrared CO (IR-CO) radii~\citep{salyk11}; the dashed horizontal line represents a one-to-one relation. With one exception, $R_{in}$(H$_{2}$) and $R_{in}$(CO) are the same to within a factor of three. The agreement between the two molecules is rather remarkable given that we are comparing different species, excited by different mechanisms (photo-excitation vs. collisional excitation), observed at different epochs in different wavebands. The notable exception is SU Aur, whose $R_{in}$(H$_{2}$) is approximately an order magnitude larger than its corresponding $R_{in}$(CO), possibly due to H$_{2}$ emission from nebulosity associated with this star. The agreement between the IR-CO emission and the UV-H$_{2}$ emission is also interesting in light of the recent discovery of large amounts of CO emission in the UV spectra of CTTSs (UV-CO; France et al. 2011b). \citet{schindhelm11} present an initial survey of this emission, showing that the line-widths of UV-CO are systematically narrower than those of UV-H$_{2}$. Their interpretation favors a picture where the UV-CO originates in a cooler molecular layer ($T_{rot}$(CO) ~$\sim$~500 K) at larger semi-major axes ($a$~$\gtrsim$~2 AU) than both the UV-H$_{2}$ and the IR-CO, consistent with the results presented here. 

\subsection{H$_{2}$ in Transitional Disks}

The majority of H$_{2}$ emission in the targets in our sample originates from $a$~$\lesssim$~3 AU. A notable outlier from the average H$_{2}$ radii presented in Table 3 is CS Cha, which lies at significantly larger $\langle$$R_{H2}$$\rangle$$_{[1,7]}$ than the rest of the disks studied here. The [1,7] line-width of CS Cha is 18~$\pm$~7 km s$^{-1}$, which makes it the only unresolved moderate inclination target\footnote{The [1,4] line-width is 29~$\pm$~12 km s$^{-1}$, consistent with the [1,7] result. Bary et al. (2008) report a near-IR line-width of 12.6 km s$^{-1}$, which would be unresolved in our COS observations.}, and allows us to place a lower limit on the average H$_{2}$ emission radius, $\langle$$R_{H2}$$\rangle$$_{[1,7]}$ $\geq$~9 AU. CS Cha is a transitional disk, showing the largest mid-IR spectral slope in our sample, $n_{13-31}$ = 2.89 (Furlan et al. 2009; see Table 3).~\nocite{furlan09} Modeling of the $Spitzer$-IRS mid-IR spectrum of CS Cha reveals a truncation of the inner disk dust distribution at $a$~$\approx$~43 AU~\citep{espaillat07a}, possibly the result of a dynamical interaction with a companion star (Guenther et al. 2007, but see also Espaillat et al. 2011).~\nocite{guenther07,espaillat11} 
The new $HST$ data presented here confirms the presence of molecular gas in the system~\citep{bary08}, and while we cannot rule out the possibility that this gas is co-spatial with the dust at $a$~$\gtrsim$~40 AU, this would imply that hot molecular gas ($T$(H$_{2}$)~$\geq$~2000 K) exists at large radial distances from the star. 
Therefore, because only ($N$(H$_{2}$)~$\gtrsim$~10$^{18}$ cm$^{-2}$) is required to produce detectable fluorescence, this emission may originate in the tenuous molecular material in the disk gap, or if Ly$\alpha$ photons can propagate through the gap, this emission may arise from the edge of the directly exposed wall at $\sim$~43 AU. 

The H$_{2}$ emitting material we observe in CS Cha may be physically associated with the [\ion{Ne}{2}] emission observed by~\citet{espaillat07a}, but that cannot be conclusively determined from the available observations.
We can rule out an origin in the optically thin dust disk inside of 1 AU, as our limit on the H$_{2}$ inner disk radius in CS Cha is $R_{in}$(H$_{2}$)~$\geq$~3 AU. 

CS Cha is one example of a generic property of our sample: H$_{2}$ is common in the inner regions of accreting transitional disks ($n_{13-31}$~$>$~0.5). In addition to CS Cha, our sample includes the well-studied transitional systems DM Tau, GM Aur, UX Tau A, LkCa15, HD 135344B and TW Hya~\citep{calvet05,calvet02,espaillat07b}. H$_{2}$ emission is found to originate inside the dust hole in these systems, $\langle$$R_{H2}$$\rangle$ $<$ $R_{dust}$, consistent with the origin of the IR-CO gas in the inner regions of transitional disks~\citep{salyk09,salyk11} and previous observations of near-IR rovibrational emission from H$_{2}$~\citep{bary08}. 

\section{Summary}

We have presented the most sensitive survey of H$_{2}$ in protoplanetary environments to date. 
The majority of this work was made possible by the combination of large effective area and low instrumental background at moderate spectral resolution provided by the $HST$-Cosmic Origins Spectrograph. We have used this survey to measure the time evolution of both the spatial distribution and the luminosity of H$_{2}$ in young, low-mass disks for the first time. Below we summarize the primary results of this work: 

\begin{enumerate}

\item We obtained far-UV spectra of 34 T Tauri stars: 27 accreting CTTSs and 7 non-accreting WTTSs. 
Of these, 100\% of the accreting sources display a measureable amount of H$_{2}$ emission, providing direct evidence for the interaction of a strong Ly$\alpha$ radiation field with the molecular disk surface. 

\item We found that the H$_{2}$ luminosity is well correlated with the Ly$\alpha$ and \ion{C}{4} luminosity, consistent with a scenario where gas-rich disks fuel larger accretion rates that produce energetic radiation.
\item The H$_{2}$ luminosity is observed to decline with age, although H$_{2}$-rich systems persist to ages $\sim$~10 Myr. 

\item We measured resolved H$_{2}$ line profiles of 23 targets with inclination angles $> $~15\arcdeg\ and 
found that these line profiles are reasonably well fit by a single Gaussian component at or near the stellar radial velocity. Assuming a disk origin for these targets, we used the line-widths to constrain the spatial distribution of the emitting molecules to $a$~$\lesssim$~3 AU in most cases. 

\item The inner radii of H$_{2}$ disks are roughly consistent with those of CO disks (measured from $\lambda$~$\sim$~5~$\mu$m spectroscopy). 

\item Strong H$_{2}$ emission is observed at $a$~$\gtrsim$~0.2~AU in a subsample of transitional disks ($n_{13-31}$~$>$~0.5). 

\end{enumerate} 

\acknowledgments

We thank Tom Ayres for custom processing of the STIS observations and the DAO of Tau team for enjoyable discussions during the course of this work. KF thanks Phil Armitage for helpful discussion regarding gas giant migration. We acknowledge the technical efforts of Nico Nell and David Morris, and ES and KF thank Brian Wood for input on Ly$\alpha$ profile reconstruction. RDA acknowledges support from the Science \& Technology Facilities Council (STFC) through an Advanced Fellowship (ST/G00711X/1). This work was supported by NASA grants NNX08AC146 and NAS5-98043 to the University of Colorado at Boulder ($HST$ programs 11533 and 12036) and made use of data from $HST$ GO programs 8041, 11616, 11828, and 12361.

\bibliography{ms}


\end{document}